\documentclass{article}
\usepackage{jheppub} 
\usepackage[utf8]{inputenc}
\usepackage{hyperref}
\usepackage{amsmath}
\usepackage{graphicx}
\usepackage{subcaption}
\usepackage{appendix}
\usepackage{booktabs}
\usepackage{placeins}

\usepackage{hyperref}
\usepackage{fancyvrb}
\usepackage{tabularx}
\usepackage{mdframed}
\usepackage{booktabs}
\usepackage{slashed}
\usepackage{units}
\usepackage{caption}
\usepackage{subcaption}
\usepackage{textcomp}
\usepackage[normalem]{ulem}
\usepackage[utf8]{inputenc}
\usepackage[usenames,dvipsnames]{xcolor}

\newcommand{\userinputcolor}{\color{gray}}

\newenvironment{bigtextfile}[1]
  { 
\mdfsetup{
    innertopmargin=1pt,
    leftmargin=1cm,
    nobreak=false,
    roundcorner=3pt,
    skipabove=3pt,
    frametitleaboveskip=-\ht\strutbox,
    frametitlealignment={\hspace*{0.010\linewidth}}
    }
  \begin{mdframed} \vspace{-0.5\ht\strutbox} \colorbox{white}{\space \scriptsize \texttt{#1}\space}\scriptsize
  } 
  {\end{mdframed}
\normalsize
}

\newenvironment{textfile}[1]
  { 
\scriptsize \mdfsetup{
    frametitle={\colorbox{white}{\space \texttt{#1}\space}},
    innertopmargin=0pt,
    leftmargin=1cm,
    frametitleaboveskip=-\ht\strutbox,
    nobreak=true,
    roundcorner=3pt,
    frametitlealignment={\hspace*{0.010\linewidth}},
    }
  \begin{mdframed}%
  }
  {\end{mdframed}

\normalsize
}

\author[a]{Nishita Desai}
\author[b]{, Florian Domingo}
\author[c]{, Jong Soo Kim}
\author[d]{, Roberto Ruiz de Austri Bazan}
\author[e]{, Krzysztof Rolbiecki}
\author[f]{, Mangesh Sonawane}   
\author[gh]{, Zeren Simon Wang}

\affiliation[a]{Department of Theoretical Physics, Tata Institute of Fundamental Research, Mumbai, India 400005 }
\affiliation[b]{Bethe Center for Theoretical Physics \&
Physikalisches Institut der Universit\"at Bonn,\\
Nu\ss allee 12, D--53115 Bonn, Germany}
\affiliation[c]{National Institute for Theoretical Physics and\\School of Physics, University of the Witwatersrand, Johannesburg, South Africa}
\affiliation[d]{Instituto de F\'isica Corpuscular, IFIC-UV/CSIC, Valencia, Spain}
\affiliation[e]{Faculty of Physics, University of Warsaw, Poland}
\affiliation[f]{Institute for High Energy Physics, Austrian Academy of Sciences, Austria}
\affiliation[g]{Department of Physics, National Tsing Hua University, Hsinchu 300, Taiwan}
\affiliation[h]{Asia Pacific Center for Theoretical Physics (APCTP)
- Headquarters San 31,\\ Hyoja-dong, Nam-gu, Pohang 790-784, Korea}

\emailAdd{nishita.desai@tifr.res.in}
\emailAdd{florian.domingo@csic.es}
\emailAdd{jongsoo.kim@tu-dortmund.de}
\emailAdd{rruiz@ific.uv.es}
\emailAdd{krolb@fuw.edu.pl}
\emailAdd{mangesh.sonawane@oeaw.ac.at}
\emailAdd{wzs@mx.nthu.edu.tw}

\newcommand{\loose}{{\itshape loose }}
\newcommand{\tight}{{\itshape tight }}

\DeclareUnicodeCharacter{2212}{-}
\title{Constraining electroweak and strongly charged long-lived particles with CheckMATE}

\abstract{Long-lived particles have become a new frontier in the exploration of physics beyond the Standard Model.  In this paper, we present the implementation of four types of long-lived particle searches, viz.\ displaced leptons, disappearing track, displaced vertex with either muons or with missing transverse energy, and heavy charged tracks. These four categories cover the signatures of a large range of physics models.  We illustrate their potential for exclusion and discuss their mutual overlaps in mass-lifetime space for two simple phenomenological models involving either a $U(1)$-charged or a coloured scalar.  
}
\preprint{\parbox[t]{3.5cm}{APCTP Pre2020-029 \\  BONN-TH-2021-02 \\ TIFR/TH/20-41}}

\begin{document}
\maketitle

\flushbottom

\section{Introduction}
 
The Large Hadron Collider (LHC) started operations over a decade ago and a large number of searches for physics beyond the Standard Model (BSM) have been performed.  In particular, the general-purpose experiments ATLAS and CMS have published over a thousand papers each, minutely testing the predictions of the Standard Model and of more exotic theories. For the simplest cases of new physics consisting of new strongly charged particles, corresponding searches already place limits on the BSM masses at about 2-3 TeV \cite{Aad:2020aze}, closing in on the maximum reach achievable at a 13 TeV proton collider.  We turn our attention therefore to the equally well-motivated but experimentally more challenging cases of new physics where BSM particles are long-lived.  A long lifetime may be the natural consequence of a compressed phase space (e.g.\ in particular dark matter models \cite{Ibe:2006de,Cirelli:2005uq,FileviezPerez:2008bj,Cirelli:2009uv,Buckley:2009kv,Johansen:2010ac,Mahbubani:2017gjh,Khoze:2017ixx}), a suppressed connection to light SM decay products caused by heavy mediators \cite{Giudice:2004tc}, or direct but feeble couplings to SM particles (possibly resulting from an approximate symmetry) \cite{Hall:2009bx,ArkaniHamed:2008qp,Graham:2012th,Zwane:2015bra,Belanger:2018sti}.  These new kinds of scenarios have raised considerable interest in the community~\cite{Alimena:2019zri} and making the results of associated searches available to reinterpretation in terms of other theory models has been a frequent request to the {\tt CheckMATE} collaboration.
 
With the end of Run 2 of the LHC, both ATLAS and CMS have turned their attention to searches for exotic physics based on new particles with long lifetimes. The signatures of a long-lived  particle (LLP) depend on its charges as well as its  decay modes and actual lifetime, and they can be fairly complicated to systematically characterize.  Several possibilities for LLP signatures indeed exist, based on the LLP decay modes:  \begin{itemize}
    \item neutral LLP $\rightarrow$ invisible or neutral stable particles $\Rightarrow$ missing momentum;
    \item neutral LLP $\rightarrow$ charged leptons $\Rightarrow$ leptons with large impact parameter (i.e. ``displaced'' leptons);
    \item neutral LLP $\rightarrow$ coloured particles $\Rightarrow$ displaced vertices, or emerging jets;
    \item stable, charged LLP $\Rightarrow$ charged track (with its time of flight dependent on mass and boost);
    \item charged LLP $\rightarrow$ invisible $\Rightarrow$ ``disappearing'' track;
    \item charged LLP $\rightarrow$ other charged stable object(s) $\Rightarrow$ kink-track or displaced vertex. 
\end{itemize}  
There is a built-in complementarity in different searches simply because particle decay follows an exponential distribution. For example, charged LLPs with intermediate lifetimes will be visible in both disappearing track and heavy charged track searches.  Similarly, a neutral particle decaying into quarks mostly in the electromagnetic or hadronic calorimeter will also likely appear as a smaller, simultaneous signal in the displaced vertex (decay in the tracker) and emerging jet (decay in the hadronic calorimeter) searches.  
 Furthermore, the lifetime of a particle in the lab-frame also depends on its boost, which means that the production mechanism can also significantly alter where in the detector the particle decays.  The same particle may accordingly result in different signal distributions depending on whether it is produced ``directly'' or in the decays of a much heavier particle (i.e.\ with a higher boost).  Finally, several decay modes may be open to a single new particle resulting in sensitivity in multiple searches.  The identification of the underlying physics therefore requires a full coverage in terms of the lifetime of new particles.  

As we can see, the identification of an LLP is highly complicated and so far, there are no standard algorithms like those available for the identification of standard objects, such as leptons, $b$- or $\tau$-tagged jets, etc. Consequently, it is not always clear how the results of a dedicated LLP search can be  ``reinterpreted'' for a physics model that differs from the tested one, though it displays a priori similar signatures.  A detailed study of models capable of LLP signatures, the reinterpretation struggles and recommendations have been detailed in the community study~\cite{Alimena:2019zri}.   In this present work, we use the signal efficiencies published by the experiments  in order to implement five searches in the {\tt CheckMATE} reinterpretation package.  The current searches use the existing respective detector implementations for ATLAS and CMS experiments in Delphes.  It should also be possible to implement dedicated searches from experiments like FASER~\cite{Feng:2017uoz}, CODEX-b~\cite{Gligorov:2017nwh} or even proposed experiments like MATHUSLA~\cite{Chou:2016lxi} if a corresponding \textsc{Delphes} module or efficiency parametrisations become available. 

{\tt CheckMATE} \cite{Drees:2013wra,Dercks:2016npn} is a public tool that allows the reinterpretation of a wide variety of ATLAS and CMS results for new physics models in a coherent and cohesive manner. It consists of an engine written in C++ that runs each analysis cut-by-cut in order to assess the final number of expected events satisfying the requirements of the corresponding analysis.  The engine is also capable of using external libraries like \textsc{Madgraph}~\cite{Alwall:2014hca} and \textsc{Pythia~8}~\cite{Sj_strand_2015}\footnote{Internal running of Monte Carlo processes or use of \textsc{Madgraph} interface in {\tt CheckMATE} is currently compatible only with the 8.2 series of \textsc{Pythia~8}.} in order to generate events, while the detector simulation is performed by  \textsc{Delphes}~\cite{de_Favereau_2014}. The User Interface and the statistical analyses are provided by a collection of Python scripts (including the \texttt{AnalysisManager}~\cite{Kim:2015wza} that guides the users through the implementation of their own analyses).

In section~2, we briefly summarize the main ingredients of the LLP recast, referring the reader to the appendix for a more complete description. Then, in section~3, we illustrate the performance of the implemented searches in two simple models with LLPs, and discuss their complementarity. Conclusions and a brief outlook are proposed in section~4. The appendix consists of a short guide for the user, as well as a more detailed presentation of the implemented LLP searches.



\section{Implementation of long-lived particle searches}

 Below we offer a brief description of the implemented LLP searches and a comparison with experimentally published results.  This includes the 8 and 13 TeV versions of the CMS displaced lepton search \cite{Khachatryan_2015,CMS-PAS-EXO-16-022}, two different displaced vertex searches \cite{Aaboud:2017iio,Aad:2020srt}, the 13 TeV ATLAS disappearing track \cite{Aaboud:2017mpt} and heavy charged particle track \cite{Aaboud:2019trc} searches.  Together, these searches are capable of probing a wide range of parameter space. Details of the implementation are available in the appendix.  Technically, each analysis is encapsulated in a detector-specific ``analysis handler'' class which provides special functions and efficiencies specific to the detector in question.  We deliberately separate the analysis handlers for long-lived particle searches from those used in prompt searches: this accounts for the fact that the implemented prompt searches do not in fact use any decay length information and all particles denoted stable by the Monte Carlo generator are then clustered into jets based on their kinematics only.\footnote{For this reason, we advise user discretion when applying prompt search limits to models with LLPs.  It may be possible to find a conservative limit from prompt searches by e.g.\ removing decay products of LLPs from the Monte Carlo events beforehand. However each case needs to be evaluated separately and we do not provide a built-in solution for that reason.}

\subsection{Displaced Lepton searches \label{sec:DLS}}

The displaced lepton searches \cite{Khachatryan_2015,CMS-PAS-EXO-16-022} look for two high-$p_T$, isolated leptons ($\ell$) with large impact parameter relative to the primary vertex.  The benchmark used for this search is motivated by R-parity violating (RPV) supersymmetry \cite{Barbier:2004ez,Graham:2012th} where a top-squark ($\tilde{t}_1$) decays via the lepton-number-violating LQD operator as $\tilde t_1 \rightarrow \ell b$.  The leptons thus produced have large $p_T$ and are well isolated.  The two searches implemented here correspond to 8 TeV \cite{Khachatryan_2015} and 13 TeV \cite{CMS-PAS-EXO-16-022} versions of the CMS displaced supersymmetry search.  The identification and fiducial acceptances are provided on generator-level events.  We therefore reproduce the Monte-Carlo production process for validation of the search. Corresponding details are provided in appendix~\ref{app:A}.

\begin{figure}[t]
\centering
\begin{subfigure}{.49\linewidth}
  \includegraphics[width=\textwidth]{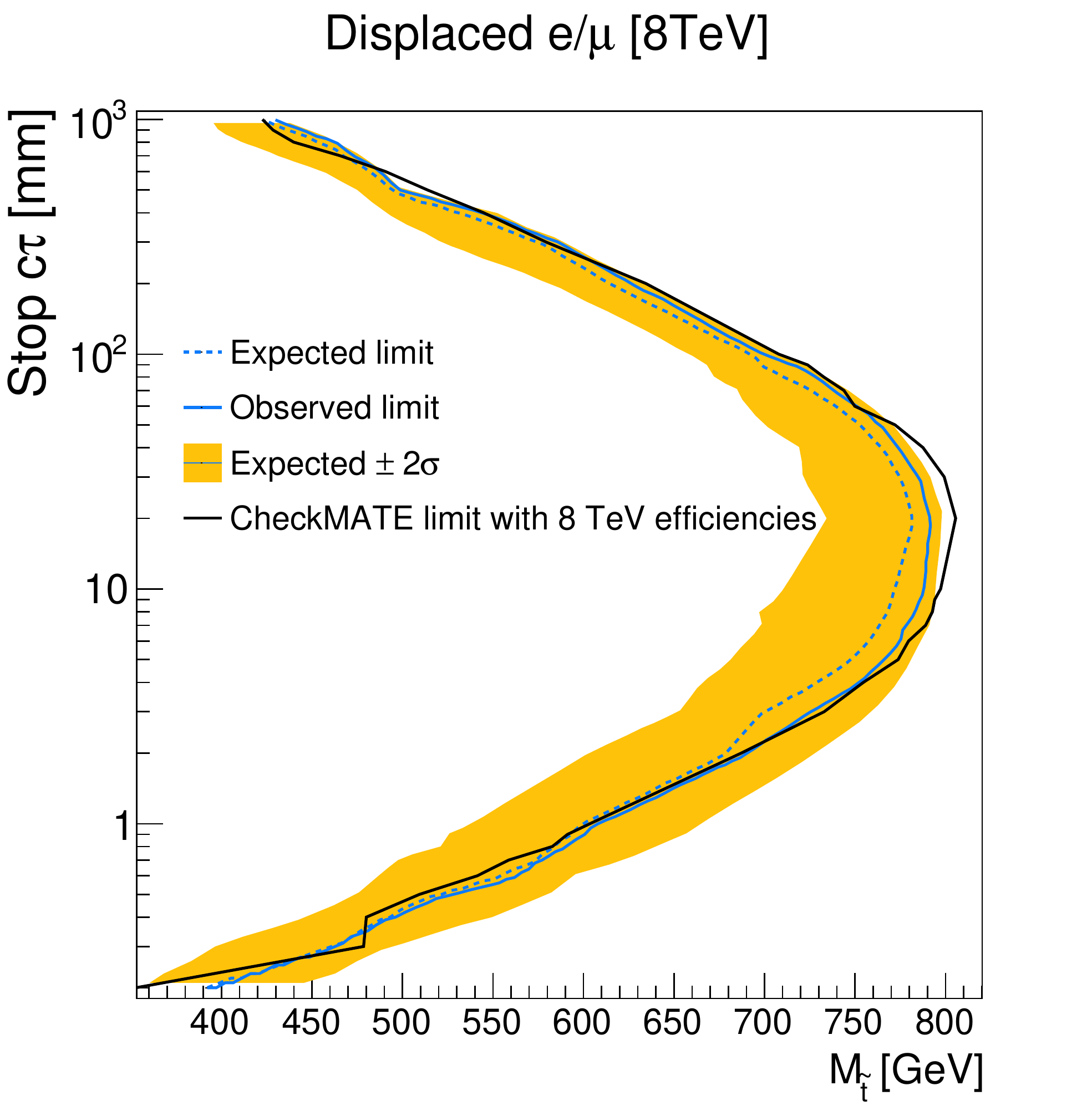}
  \caption{8 TeV}
  \label{fig:8TeVDLex}
\end{subfigure}
\begin{subfigure}{.49\linewidth}
  \includegraphics[width=\textwidth]{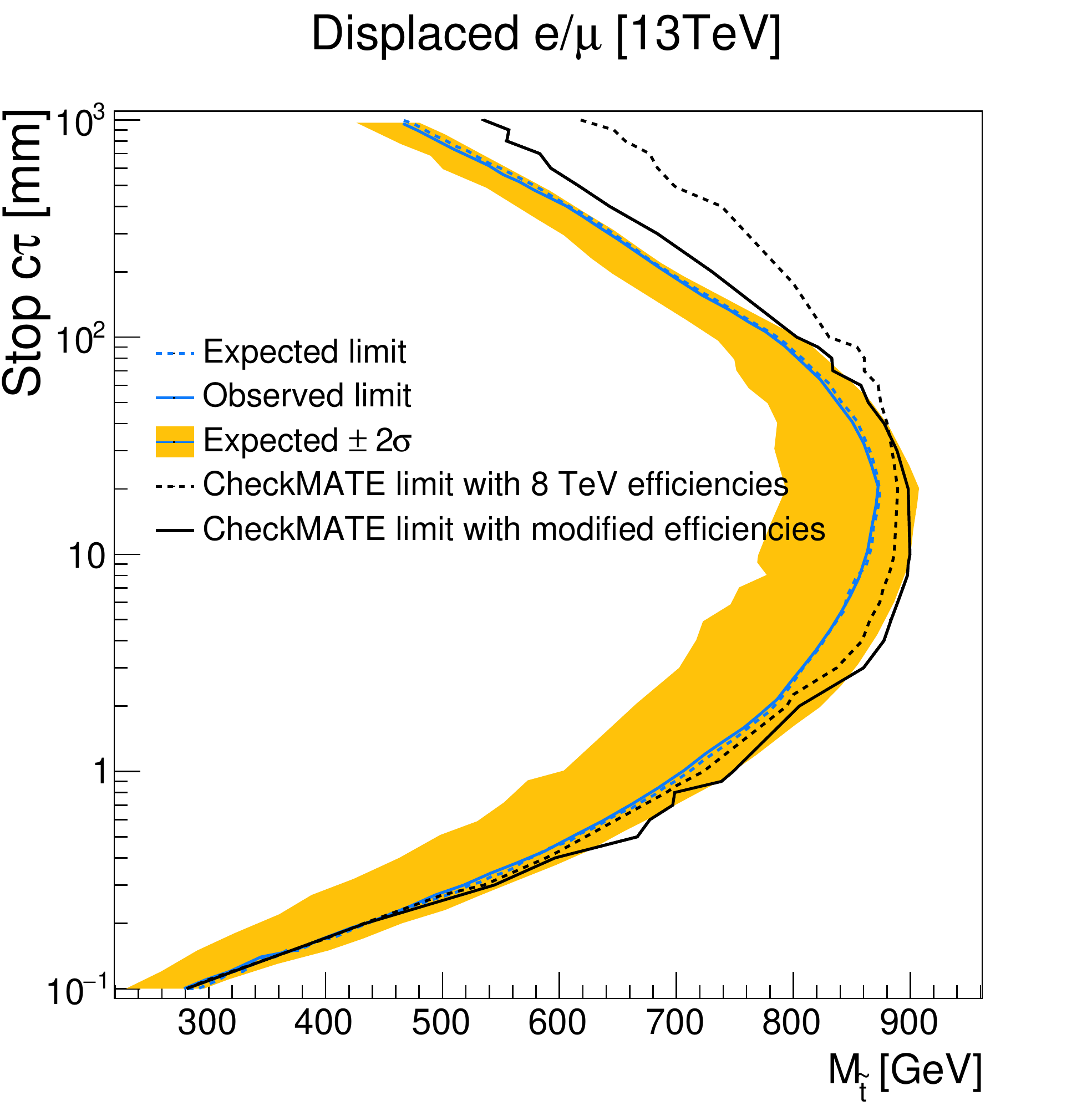}
  \caption{13 TeV}
  \label{fig:13TeVDLex}
\end{subfigure}
\caption{A Comparison of the exclusion limits on the Displaced Lepton search provided by CMS  with those obtained from {\tt CheckMATE} (left: 8 TeV, 19.7 fb$^{-1}$; right: 13 TeV, 2.6 fb$^{-1}$). }
\label{fig:DLexclusion}
\end{figure}

The event selection for both 8 TeV and 13 TeV analyses was performed in two stages.  The first stage (i.e.\ preselection) selects events with exactly one electron and one muon with opposite electric charges, each expected from the decay of a different top squark.  Further selection cuts and isolation requirements are then applied.  In the second stage, the events are classified into three signal regions (SR) corresponding to increasing ranges of the leptonic impact parameter $d_0$.

The validation results are shown in Fig.~\ref{fig:DLexclusion}, with the experimental exclusion limits displayed in blue, while the recast produces the black exclusion bound. A reasonable agreement is observed at $8$~TeV in Fig.~\ref{fig:8TeVDLex}. The situation at $13$~TeV is somewhat more subtle.

Indeed, as efficiencies have been provided by the experimental collaboration for the $8$~TeV, but not for the $13$~TeV search---in particular, the considered ranges of $p_T$ and $d_0$ do not match and the modelling of efficiencies in this latter case appears as an important assumption in the recast. An attempt to validate the 13 TeV search using the 8 TeV efficiencies results in poor agreement with the numbers in the signal region at large impact parameters, and therefore in a much stronger expected limit for high values of the LLP lifetime $c\tau$ (see the dashed black curve in Fig.~\ref{fig:13TeVDLex}).  A simple linear interpolation performs rather poorly as well.  We therefore make a conservative estimate of this detector effect by adding a single bin for the $d_0$ range (20 mm - 100 mm) and determine the associated efficiency via a $\chi^2$-fit of the expected number of events in all three signal regions.  The outcome of this procedure is displayed in Fig.~\ref{fig:13TeVDLex} as a solid black line and shows a considerably improved agreement  with the expected exclusion limits.   Exact numbers in each of the signal regions are produced in Appendix~\ref{app:A}. 
\FloatBarrier

\subsection{Displaced Vertex searches --- DV + 
MET}\label{sec:1710.04901}

This ATLAS search \cite{Aaboud:2017iio} looks for high-mass displaced vertices (DVs), reconstructed from five or more tracks. Large missing transverse momentum is also required. The outcome of $32.8~\text{fb}^{-1}$ of $13$~TeV collision data is a yield consistent with the expected background.

The template considered by the ATLAS collaboration consists in the (strong) production of a pair of long-lived gluinos ($\tilde{g}$), then decaying into light quarks and stable neutralinos. Heavy squark mediators result in suppressed gluino decay widths. Recast instructions were provided in \cite{DVrecast} and include a preselection at generator-level, followed by the application of parametrized efficiencies. Previously to our implementation, this strategy has been applied with success by the publicly available codes \cite{Cottincode,Lessacode} (see also Contribution~$22$ in \cite{Brooijmans:2018xbu}). Details on the implementation are provided in appendix~\ref{app:C}.

\begin{figure}[tbhp]
\begin{center}
\includegraphics[width=0.49\textwidth]{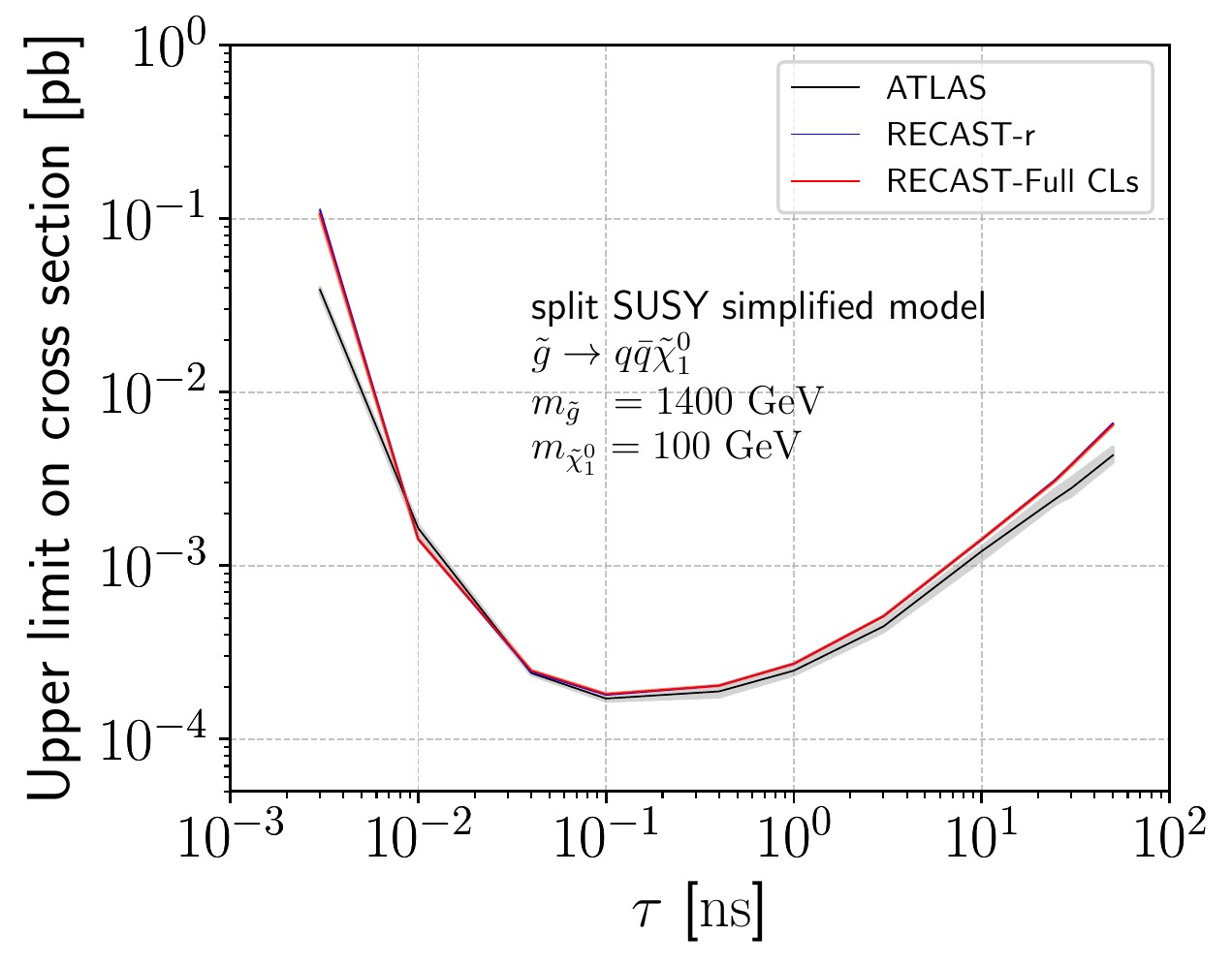}
\includegraphics[width=0.49\textwidth]{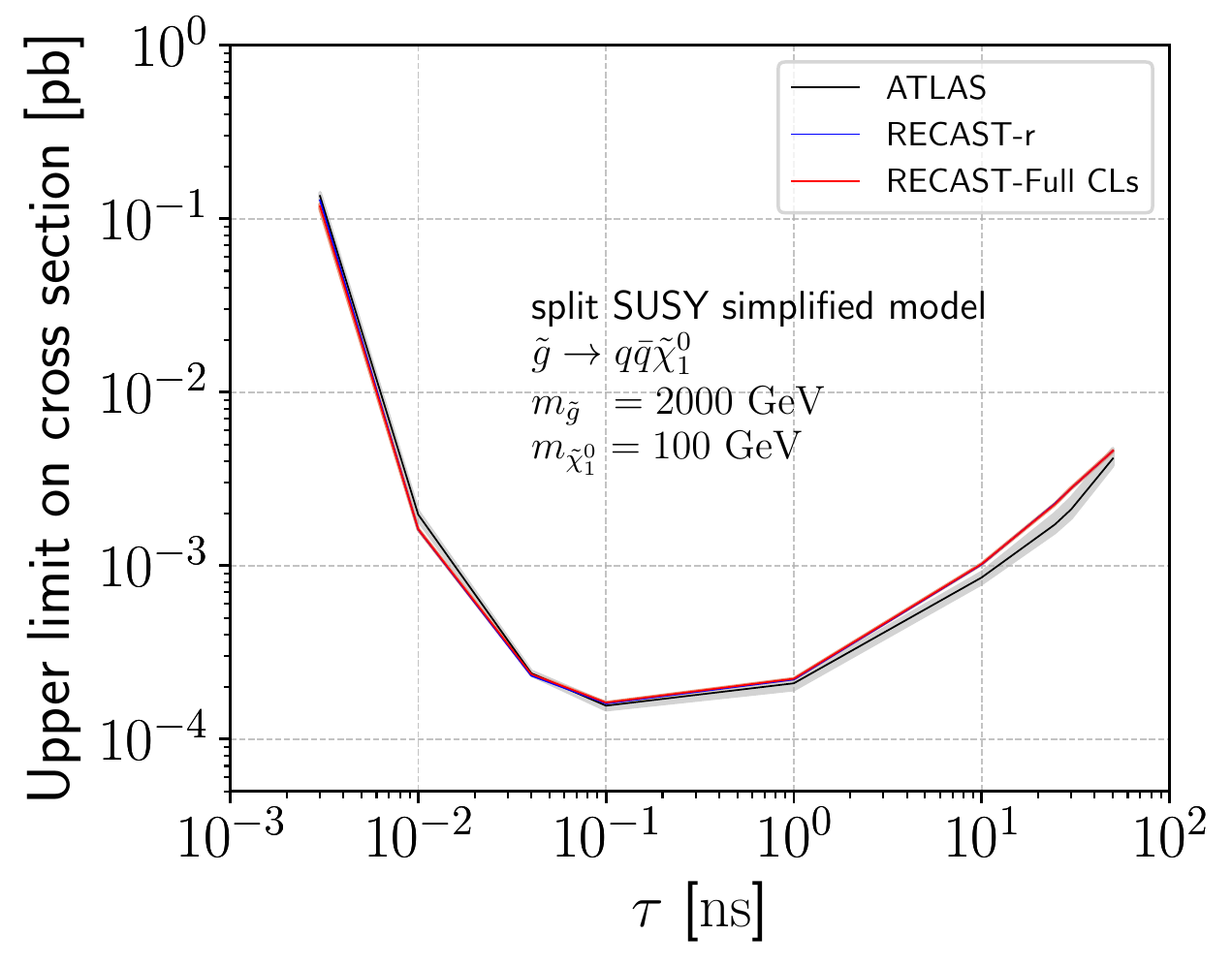}\\ \vspace{-0.2cm}
\includegraphics[width=0.49\textwidth]{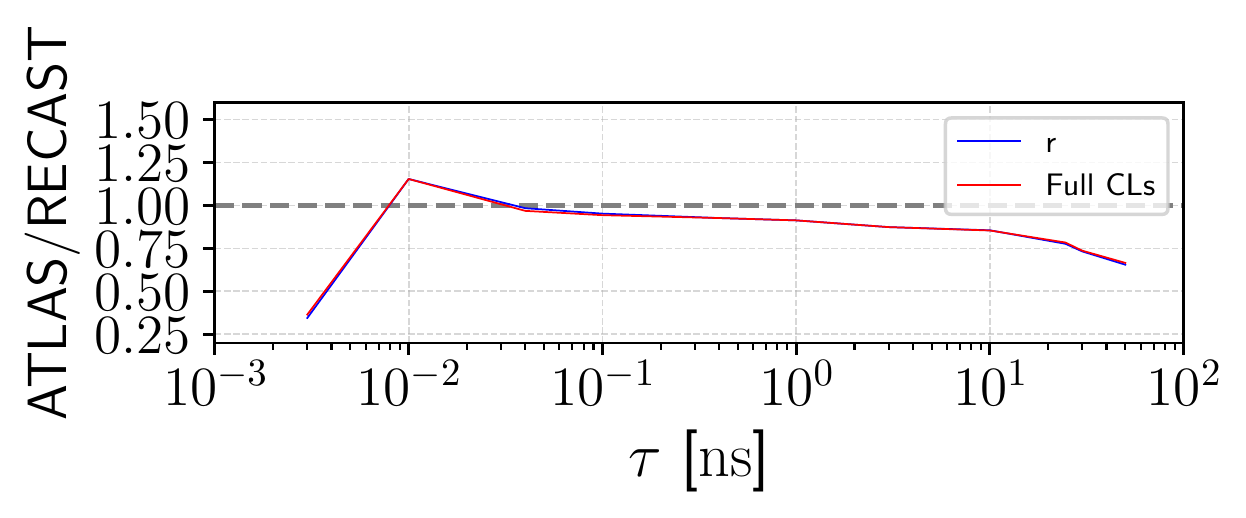}
\includegraphics[width=0.49\textwidth]{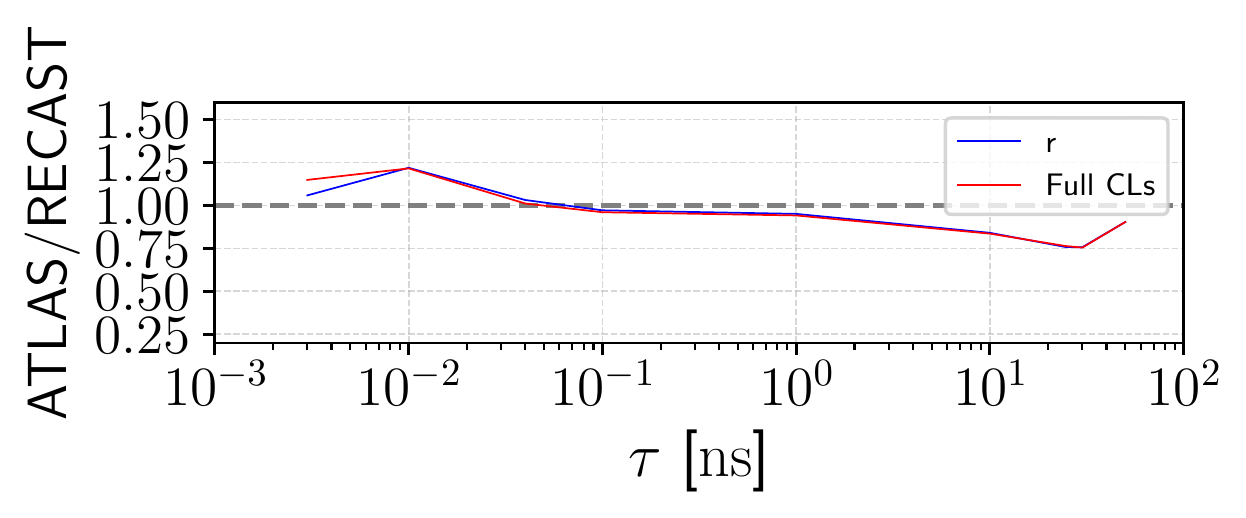}
\end{center}
\caption{Validation of the DV+MET search in the scenario with large mass-splitting for two different benchmarks (left $m_{\tilde{g}}=1.4$~TeV, right: $m_{\tilde{g}}=2$~TeV.). The bottom panel in both cases shows the ratio of the published ATLAS exclusion to the one obtained by {\tt CheckMATE}.  There is no observable difference between using the full CLs method and an exclusion based on the ratio ($r$) of cross section from {\tt  CheckMATE} of events passing all cuts to the 95\% upper limits on cross sections published by ATLAS. }
\label{fig:validation_1710_04901}
\end{figure}

For the validation, we considered the limits on the gluino production cross-sections presented in the ATLAS paper. In a first scenario, the long-lived gluino and the neutralino LLP are separated by a wide mass-gap, with the neutralino fixed at $100$~GeV while the gluino takes mass of $m_{\tilde{g}}=1.4$~TeV or $2$~TeV. The LLP lifetime is varied between $\tau_{\tilde{g}}=0.003$~ns and $50$~ns.
The results from the recast search are shown in Fig.~\ref{fig:validation_1710_04901}, using two statistical approaches: the simplified evaluation of {\tt CheckMATE} defining a ratio $r$ (blue curve)---see Eq.~(1) of \cite{Dercks:2016npn}---and the full $p$-value analysis (red curve); both return very similar limits.
The statistical uncertainty in the simulation is below percent level ($10^6$~events are generated at each point), except for the end point $\tau_{\tilde{g}}=0.003$~ns, where it reaches $3-4\%$. The $95\%$\,CL limits from the experimental analysis is shown in black.
We observe a general qualitative agreement. In the lower row of plots of Fig.~\ref{fig:validation_1710_04901}, the experimental observed limits are normalized to the limiting cross-sections of the recast procedure (with $r$-approach in blue and $p$-values in red). Quantitatively, we find that the bounds agree within $20\%$ of the cross-section value, with outliers at up to $50\%$ discrepancy for small lifetimes.

Nevertheless, this apparent success of the recast strategy with efficiencies applied on truth-level objects needs to be tempered as it seems to perform worse in the case of a compressed spectrum. This was confirmed to us by the authors of  \cite{Cottincode,Lessacode}. A more detailed comparison is provided in the appendix.

\FloatBarrier

\subsection{Displaced Vertex searches --- DV + $\mu$}\label{sec:dvplusmu}

In this section we discuss a search for massive, long-lived particles decaying to final states with a DV and an energetic muon~\cite{Aad:2020srt}. The search analyzed $139~\text{fb}^{-1}$ of data collected by ATLAS at the centre of mass energy 13~TeV.

The benchmark process considered by the experiment was pair production of top squarks followed by the RPV decay into a light quark and a muon. Other physics scenarios, for example, long-lived lepto-quarks, right-handed neutrinos or long-lived electroweakinos in RPV, could result in similar signals including a DV and a muon. In Section~\ref{sec:performance} we apply this search to sbottom pair-production followed by the RPV decay.

The event selection defines two mutually exclusive trigger-based signal regions: $E_\mathrm{T}^\text{miss}$ Trigger SR and Muon Trigger SR. The former requires significant missing transverse momentum ($>180$~GeV), while the latter is recorded with the muon trigger and has low ($<180$~GeV ) transverse momentum. Additionally, at least one displaced vertex is required to be present in the fiducial region. There is no explicit requirement for a signal muon to originate from the reconstructed vertex.  

\begin{figure}[tbh!]
\begin{center}
\includegraphics[width=0.7\textwidth]{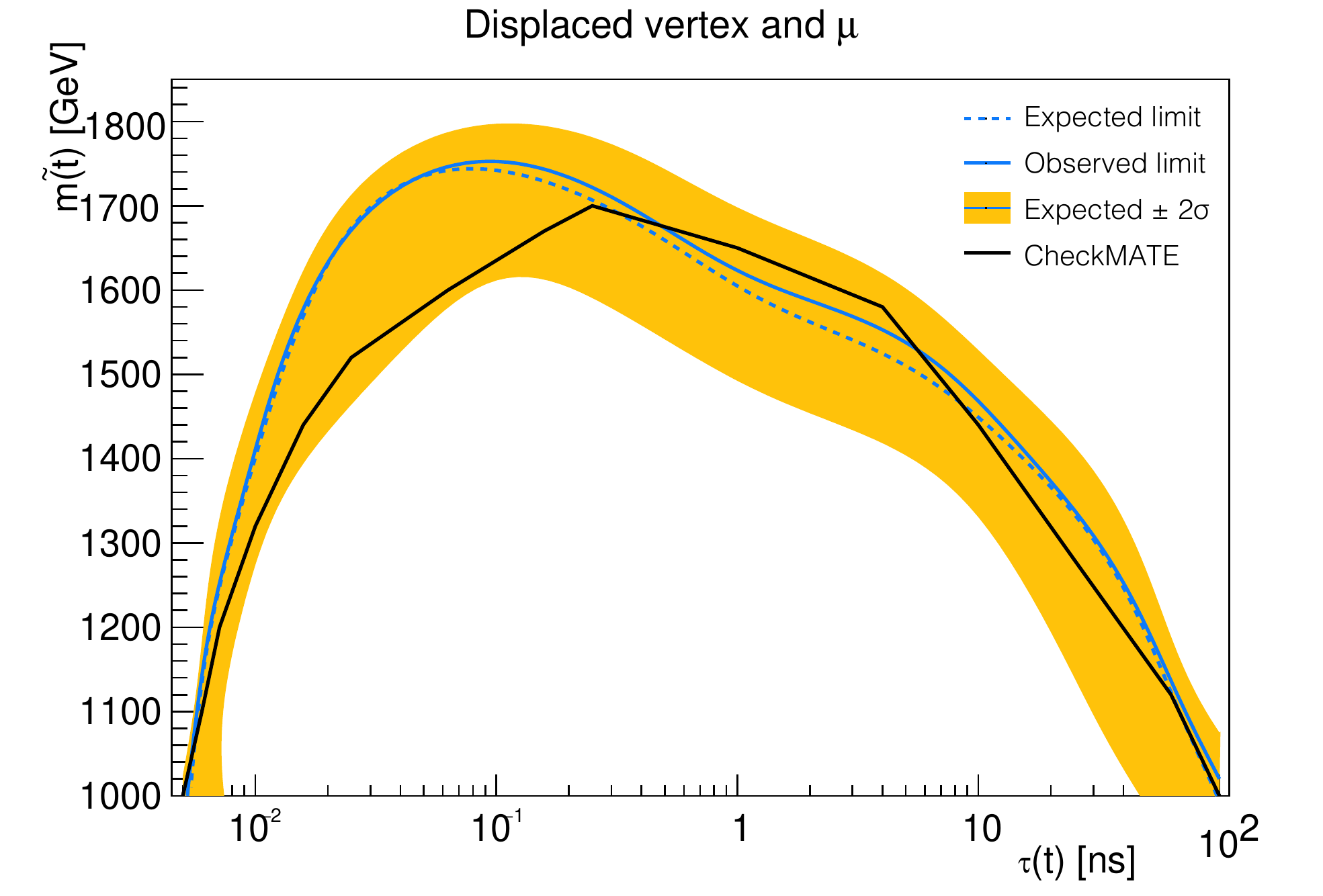}
\end{center}
\caption{Comparison of exclusion limits in the stop mass and stop lifetime plane, $m_{\tilde{t}}$--$\tau_{\tilde{t}}$, reported by ATLAS: expected -- dashed blue; observed -- solid blue and {\tt CheckMATE} -- black solid. The yellow band shows a 2-sigma range of the ATLAS expected exclusion limit. \label{fig:validation_2003.11958}}
\end{figure}

The search was validated using a benchmark RPV-supersymmetric (SUSY) model for the process $pp \to \tilde{t}_1 \tilde{t}_1$, $\tilde{t}_1 \to \mu\, q$. In Figure~\ref{fig:validation_2003.11958} we show a comparison of the ATLAS result and {\tt CheckMATE} recasting in the stop lifetime-mass plane, $\tau_{\tilde{t}}$--$m_{\tilde{t}}$. The yellow band shows a 2-sigma range of the ATLAS expected exclusion limit, the blue solid line is the ATLAS observed exclusion while the blue dashed the ATLAS expected exclusion, and the black solid line shows an exclusion line obtained with {\tt CheckMATE}. Generally a good agreement is observed, however in a range of lifetimes $0.01$--$0.1$~ns, the recast exclusion is significantly weaker, though within the 2-sigma band. Further details can be found in Appendix~\ref{app:E}.  

\FloatBarrier

\subsection{Heavy Charged Particles searches} \label{sec:1902.01636}

In this section we focus on the search for heavy charged long-lived particles performed by the ATLAS experiment using a data sample of $36.1~\text{fb}^{-1}$ of collisions at 13 TeV \cite{Aaboud:2019trc}. In our implementation we cover searches for long-lived charginos and sleptons.

The ATLAS collaboration reported no significant excess of observed data events above the expected background in this search. Thus, the collaboration have published upper limits at 95\% confidence level on the cross-sections for stau and chargino production for specific benchmark models. These limits have been obtained applying the CLs prescription \cite{Read:2002hq}.

For the validation of our implementation in  {\tt CheckMATE} we have employed \textsc{HistFitter} \cite{Baak:2014wma} to estimate the CLs while running $10^5$ toy-experiments, given the low backgrounds for the considered signal regions and assuming a 10\% signal uncertainty. 

\begin{figure}[tbh!]
\begin{center}
 \includegraphics[scale=0.45]{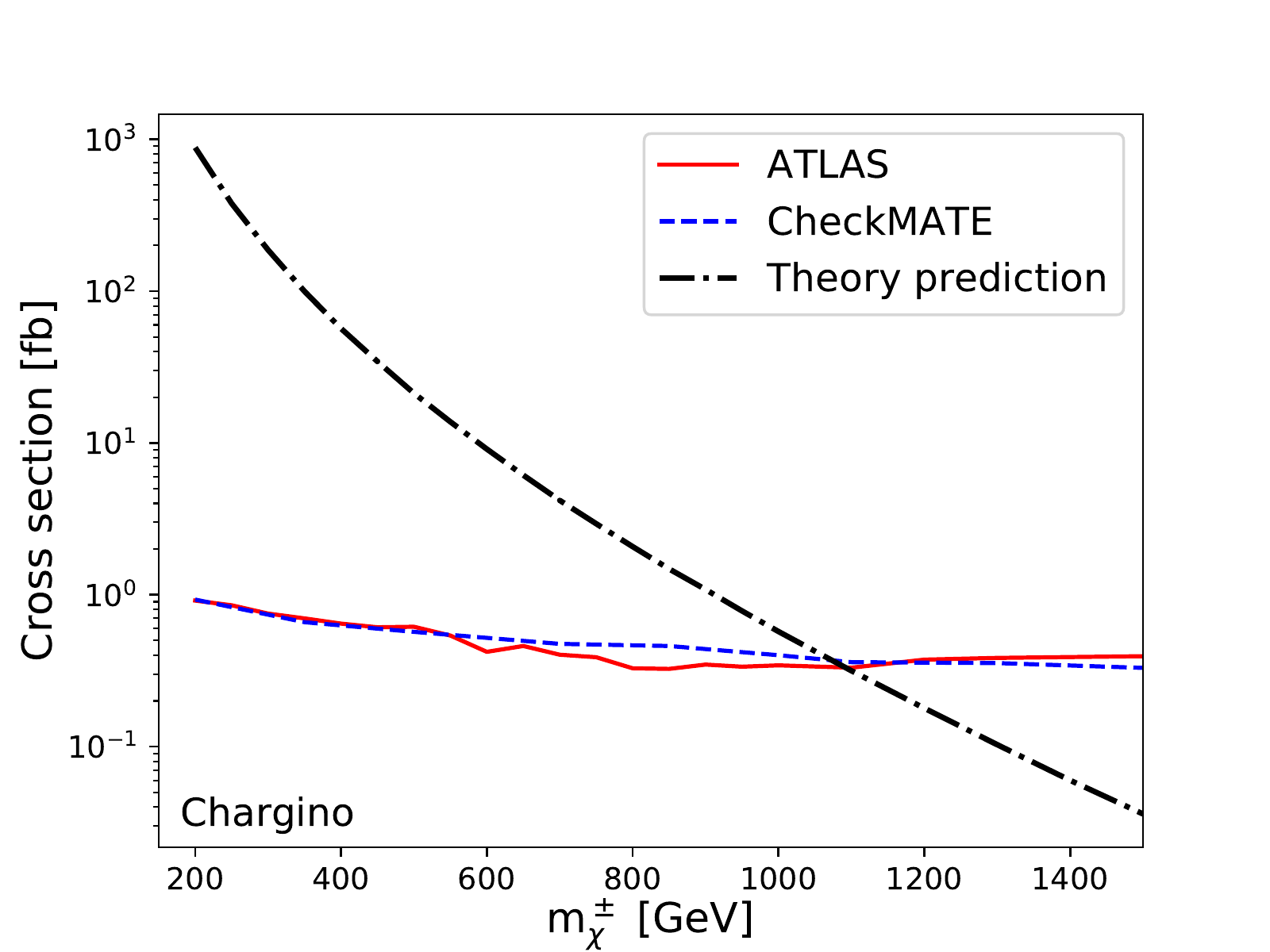}
 \includegraphics[scale=0.45]{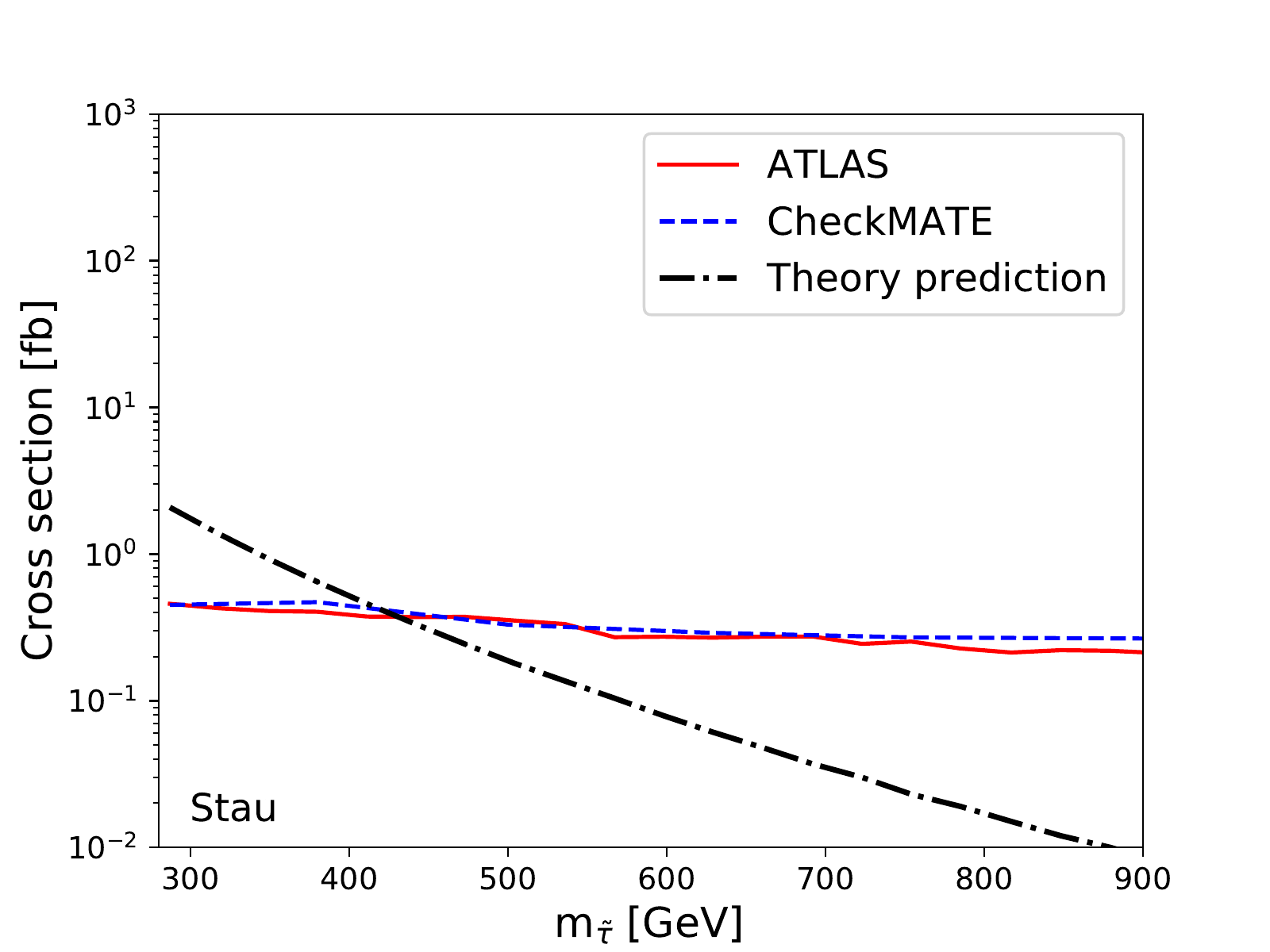}
\end{center}
	\caption{Validation of the HCP search -- observed upper cross-section limits using the best expected signal region for chargino pair production (left) and stau pair production (right). ATLAS results are shown as a solid red line while {\tt CheckMATE} predictions are in form of a dashed blue line. The theory prediction is displayed as a dot-dashed black line.}
	\label{fig:validation_1902}
\end{figure}

The validation has been performed through the comparison with the observed upper cross-section limits reported by the ATLAS collaboration, as is depicted in Fig.~\ref{fig:validation_1902}. A very good qualitative agreement is visible between the ATLAS (red) and the {\tt CheckMATE}-derived (dashed blue) limits for both chargino and stau scenarios. More details can be found in Appendix~\ref{app:heavychargedparticle}.

\FloatBarrier


\subsection{Disappearing track searches}
\label{subsect:disappearingTrack}
The ATLAS collaboration presented a search\footnote{\textsc{HEPData} accessed from \url{https://doi.org/10.17182/hepdata.78375.v5}}~\cite{Aaboud:2017mpt} for direct electroweak (EW) gaugino or gluino pair production with wino-like electroweakinos (hence near-degenerate charged and neutral $SU(2)$-triplet fermions). The chargino decays via $\tilde \chi^+ \rightarrow  \pi^+ \tilde{\chi}^0$; the neutralino is stable. The experimental collaboration analysed data based on the integrated luminosity of 36.1 fb$^{-1}$ recorded between 2015 and 2016. Due to the small mass difference between the two states (which is of the order of the pion mass), the chargino is long-lived and the decay products are entirely invisible to the detector.  Thus the chargino appears as a ``disappearing'' track, i.e.\ a track that does not reach the outer edges of the tracker detector but stops before.  In order to define a trigger isolating the signal from large SM backgrounds, the search additionally demands a large momentum jet from initial-state radiation or four jets originating from the gluino decay. The observed number of events is consistent with the SM expectations and constraints for wino-like charginos with a lifetime of 0.2 ns,  and mass up to 460 GeV are derived. In the strong production channel, where the chargino emerges from the decay of a gluino (colour octet fermion), limits on gluino masses up to 1.65 TeV are reported, under the assumption of a chargino mass of 460 GeV and lifetime of 0.2 ns.  

In order to reinterpret this search, we follow all procedures regarding production of signal events described in the original paper as closely as possible.  A detailed description can be found in appendix~\ref{app:B}.  The kinematic cuts for both signal regions are summarised in Table \ref{disappearingTrackSR}.  ATLAS further applies quality requirements, as that the tracklet is required to have hits in all four pixel layers, and a disappearance condition is demanded for each event, as the number of SCT hits associated with the tracklet must be zero. Although, the two latter cuts are impossible to simulate in a phenomenological study, ATLAS provides efficiency maps for the tracklets for EW SR and strong SR, respectively.  In addition, the collaboration provides a transverse momentum smearing function for the chargino.  We also use the benchmark SLHA files for the EW and strong scenarios, the pseudo analysis code, and all relevant data made publicly available at \textsc{HEPData} \cite{DisSearchData}.

\begin{table}[tbh]
\begin{center}
  \begin{tabular}{|l | l |}
    \hline
    EW SR & strong SR\\
    \hline
    at least one jet with $p_T>140$ GeV & $p_T(j_1)>100$ GeV, $p_T(j_2)>50$ GeV, $p_T(j_3)>50$ GeV\\
    $E_{\text{T}}^{\text{miss}}>140$ GeV & $E_{\text{T}}^{\text{miss}}>150$ GeV\\
    $\Delta\Phi(E_{\text{T}}^{\text{miss}}$, jets($p_T>50$ GeV)) $>1.0$ & $\Delta\Phi(E_{\text{T}}^{\text{miss}}$, jets($p_T>50$ GeV)) $>0.4$\\
    cuts on charged LLP with $p_T>100$ GeV & cuts on charged LLP with $p_T>100$ GeV\\
    \hline
  \end{tabular}
  \caption{Summary of the selection criteria for signal events for direct electroweakino production and the strong channel channel where the chargino is produced in gluino decays.}
  \label{disappearingTrackSR}
\end{center}
\end{table}
\begin{table}[h]
\begin{center}
  \begin{tabular}{|l | c | c | c | c | c |}
    \hline
    & CM $\tilde\chi_1^\pm\tilde\chi_1^\pm$  & CM $\tilde\chi_1^+\tilde\chi_1^0$ & CM $\tilde\chi_1^-\tilde\chi_1^0$ & CM all channels & ATLAS \\
    \hline
    \hline
    Trigger & 445.1 & 624.0 & 274,4 & 1343.5 & 1276 \\
    \hline
    Lepton Veto & 423.4 & 608.5 & 267.3 & 1308.2 & 1181 \\
    \hline
    MET and jet requirements & 164.2 & 229.6 & 101.0 & 494.8 & 579\\
    \hline
    EW SR & 5.2 & 4.4 & 1.6 & 11.2 & 13.5\\
    \hline
  \end{tabular}
  \caption{Cutflow comparison for a chargino produced in direct electroweak production with ($m_{\tilde \chi^\pm_1}$, $\tau_{\tilde \chi^\pm_1}$) = (400 GeV, 0.2 ns).}
    \label{disappearingTrackCutflowEW}
  \end{center}
\end{table}

\begin{table}[tbh!]
\begin{center}
  \begin{tabular}{| l | c | c |}
    \hline
    & CM & ATLAS \\
    \hline
    \hline
    Trigger & 289 & 285 \\
    \hline
    Lepton Veto & 277 & 278 \\
    \hline
    MET and jet requirements & 216 & 202\\
    \hline
    strong SR & 11 & 11\\
    \hline
  \end{tabular}
  \caption{Cutflow comparison for a chargino produced in strong production channel with ($m_{\tilde g}$, $m_{\tilde \chi^\pm_1}$, $\tau_{\tilde \chi^\pm_1}$) = (1600 GeV, 500 GeV, 0.2 ns) in the high--MET region.}
  \label{disappearingTrackCutflowStrong}
\end{center}
\end{table}

We use the ATLAS benchmark points as the test case scenarios which correspond to $ (m_{\tilde \chi^\pm_1}, \tau_{\tilde \chi^\pm_1}) = (400\,\rm{GeV},\, 0.2\,\rm{ns})$ for the EW signal region and $
(m_{\tilde g}, m_{\tilde \chi^\pm_1}, \tau_{\tilde \chi^\pm_1}) = (1600\,\rm{GeV},\, 500\,\rm{GeV},\, 0.2\,\rm{ns})$ in the strong signal region. The results for the EW and strong SR are summarized in Table~\ref{disappearingTrackCutflowEW} and \ref{disappearingTrackCutflowStrong}, respectively and our recast results show satisfactory agreement with the public ATLAS results.

We did not validate the disappearing track search in a grid scan with ATLAS exclusions, since the event generation requires matched events with two additional partons in the final state in order to reproduce the ATLAS cutflows. As only a few events would pass all selection cuts, such a scan would be costly to perform and we thus opted against it.
\FloatBarrier

\section{Performance and interplay in LLP scenarios\label{sec:performance}}
In this section, we consider two simple LLP scenarios that put forward the complementarity of the implemented searches.

\subsection{Electroweak LLP}

\begin{figure}
    \centering
    \includegraphics[scale=0.5]{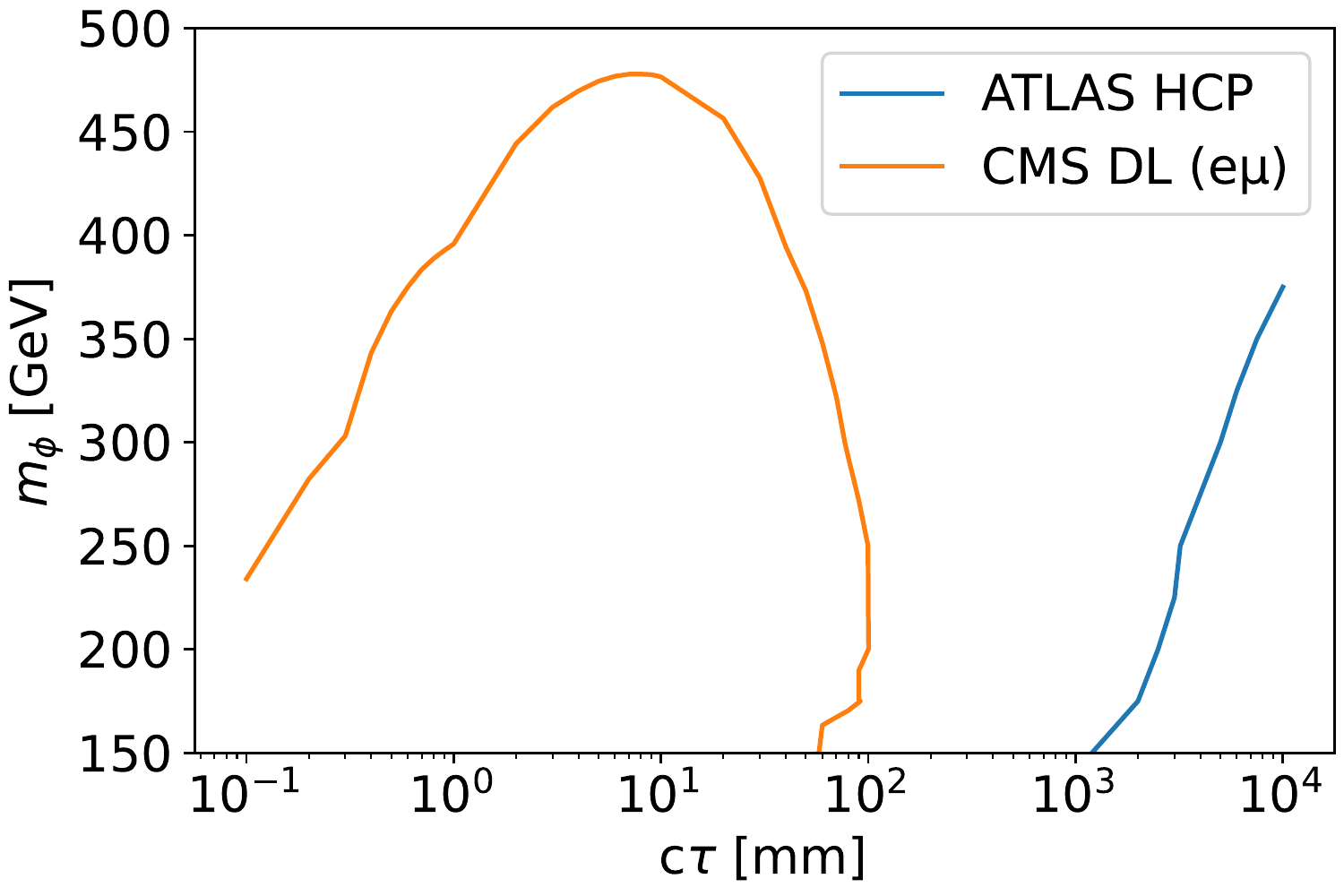}
    \includegraphics[scale=0.5]{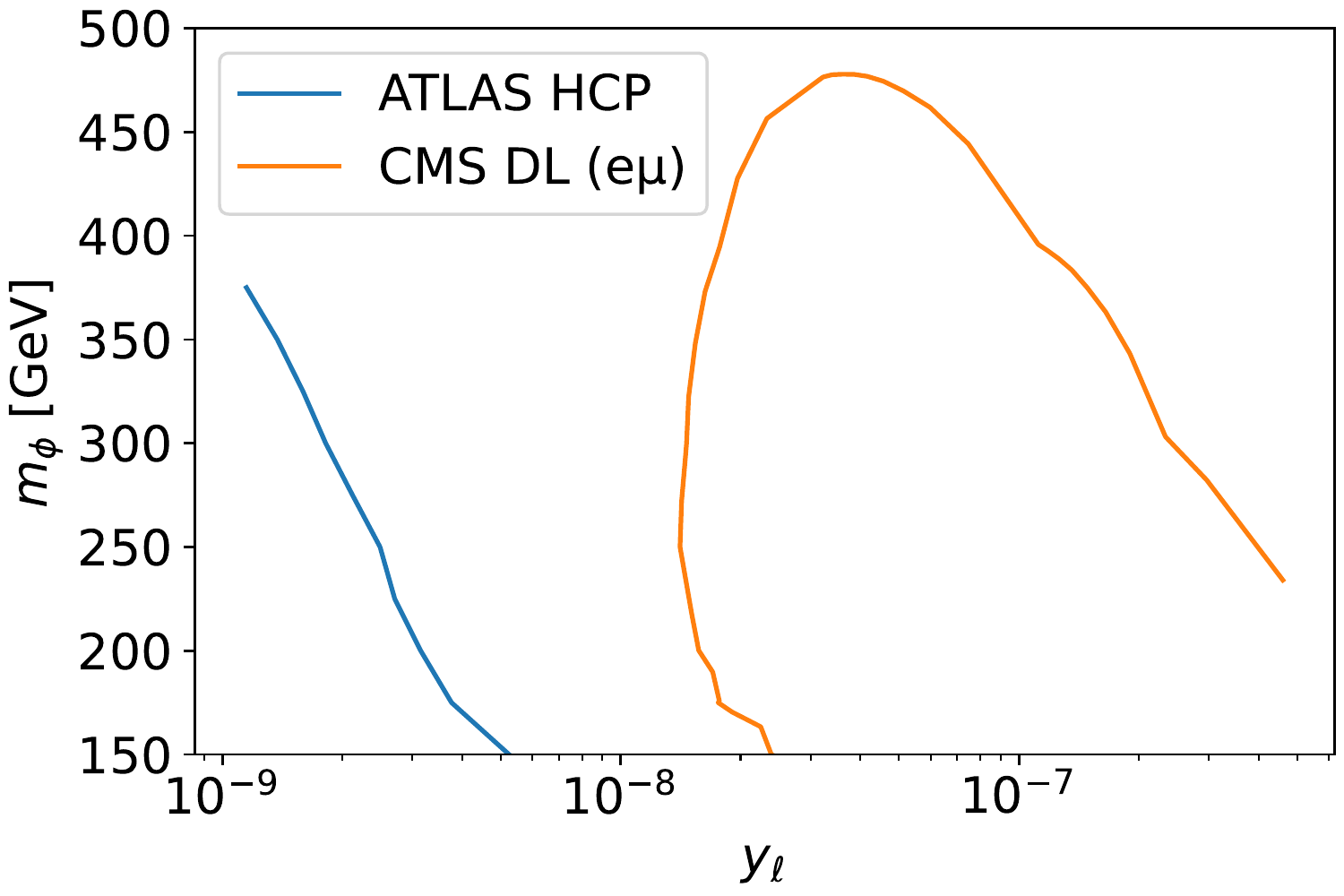}
    \caption{Upper limits in the lifetime-mass plane (left) and coupling-mass plane (right) of the LLP scalar $\phi$. We do not get limits from the disappearing track search due to a veto on leptons in the event. \label{fig:EWLLP_bounds}  
    }
\end{figure}

We first consider a model addressing the case of electroweakly produced LLPs.  It extends the SM with a scalar ($\phi$), charged under $U(1)_Y$ and sharing the gauge quantum numbers of a right-handed lepton, and a SM-singlet Dirac fermion ($\chi$).  An extra $Z_2$-symmetry under which both $\phi$ and $\chi$ are charged further constrains their interactions.  The scalar is produced in pairs via the Drell-Yan process $pp \rightarrow \phi^* \phi$. After that it decays according to $\phi \rightarrow \ell \chi$, mediated by the Yukawa coupling $y_\ell \phi \bar \ell_R \chi + \mathrm{h.c}$. The singlet fermion is assumed to be stable.  The model has been implemented in \textsc{Pythia~8}.  The Lagrangian is simply 
\begin{equation}
    \mathcal{L} = \frac{1}{2}D_\mu \phi D^\mu \phi - \frac{1}{2}m_{\phi}^2 \phi^2+ \bar \chi (i\gamma^\mu \partial _\mu) \chi - m \bar \chi \chi - \sum_\ell (y_\ell \phi \bar \chi \ell_R + \mathrm{h.c.}) \, .
\end{equation}

For small values of the Yukawa coupling, the scalar $\phi$ is long-lived and could be visible in the charged track searches.  In the case where the lifetime is too short to leave the required hits in the tracker modules, the displaced lepton search might detect the products of the decay.  To examine the complementarity of these searches, we set $y_e = y_\mu$, ensuring that the targets of the displaced lepton search are actually produced. As the lifetime is determined by the smallness of the Yukawa coupling and is fairly independent of the mass of $\chi$, we set $m_\chi = 10$ GeV.  These assumptions of course mean a focus on a very specific scenario.

The exclusion results obtained with {\tt CheckMATE} in the lifetime-mass plane are shown in Fig.~\ref{fig:EWLLP_bounds}. The CMS displaced-lepton searches appear sensitive to the considered scenario for LLP lifetimes up to $\sim1$\,ns and masses in the range $100-500$\,GeV. The Heavy Charged Particle search from ATLAS impacts the high-lifetime area $c\tau>1$\,m, in a comparable range of masses. The disappearing track search does not show any sensitivity, due mainly to cross-sections much weaker in the considered model than the targeted range of the search: the scalar production cross section is indeed suppressed compared to the fermionic one. Moreover, the original disappearing track final state consists of multiple final states (all wino final state configurations), e.g. the NLO cross section for scalar leptons of mass 125 GeV is about 0.045 pb whereas the combined wino pair production cross section is about 1~pb. Finally, the cross section is further suppressed since the disappearing track search basically triggers on monojet events which requires a hard recoil of the scalar pair against a hard jet. As a consequence, the current model cannot be probed with the disappearing track search which is focused on mass degenerate wino-like electroweakinos. Still, it is obvious that two LLP searches aiming at quite different signatures can interplay and lead to complementary exclusion bounds. Current prompt limits on pair production of scalars that decay to electron or muon with missing energy are at 250 GeV with full Run 2 data~\cite{Aad:2019qnd}.

\subsection{Strongly-interacting LLP} 
Just as EW-charged LLPs manifest in the form of track-based signatures, strongly charged LLPs result in jets originating from a secondary vertex.  A possible example is provided by the minimal supersymmetric standard model, when small R-parity violating couplings \cite{Barbier:2004ez} open up decay channels of coloured R-odd particles.  We shall focus in particular on the LQD coupling $\lambda'_{223}$ where the subscripts refer to an interaction between the second lepton doublet superfield, the second left-handed quark doublet superfield and the right-handed bottom superfield.  This results in various possible decay channels like $\tilde b_1 \rightarrow \mu c$ or $\tilde b_1 \rightarrow \nu s$, where $\tilde b_1$ is the lightest  bottom squark, which we assume to (nearly) coincide with the long-lived right-handed sbottom.

To test this scenario, we employ a simplified supersymmetric model where all the new-physics particles take a mass at the $10$~TeV scale, beyond the discovery reach of the LHC, with the exception of the lightest bottom squark of right-handed type $\tilde{b}_1$, whose mass is scanned over in the electroweak-TeV range.  The lifetime of $\tilde{b}_1$ is then entirely determined by the size of the coupling $\lambda'_{223}$ and can be varied freely. We generate samples of $10^6$ events for $pp \rightarrow \tilde{b}_1 \tilde{b}^*_1$ using \textsc{Pythia~8}. The production cross section is normalized to  the NNLO$_{\text{approx}}$ + NNLL values provided by the LHC cross-section Working Group \cite{SQprod}.

Several LLP searches are potentially sensitive to this scenario. The decay channel muon + jet is an obvious target for the DV+muon analysis, while the decay into neutrino + jet enters the scope of the DV+MET search. Finally, the bottom squark could be detected as a heavy long-lived charged particle. However, we do not compute limits from the corresponding search due to known uncertainties in the hadronization of such long-lived strongly charged scalars.
In addition, the disappearing track search is insensitive here because the displaced jets produced from the sbottom decays are hard, as a general rule.

\begin{figure}[t]
    \centering
    \includegraphics[scale=0.5]{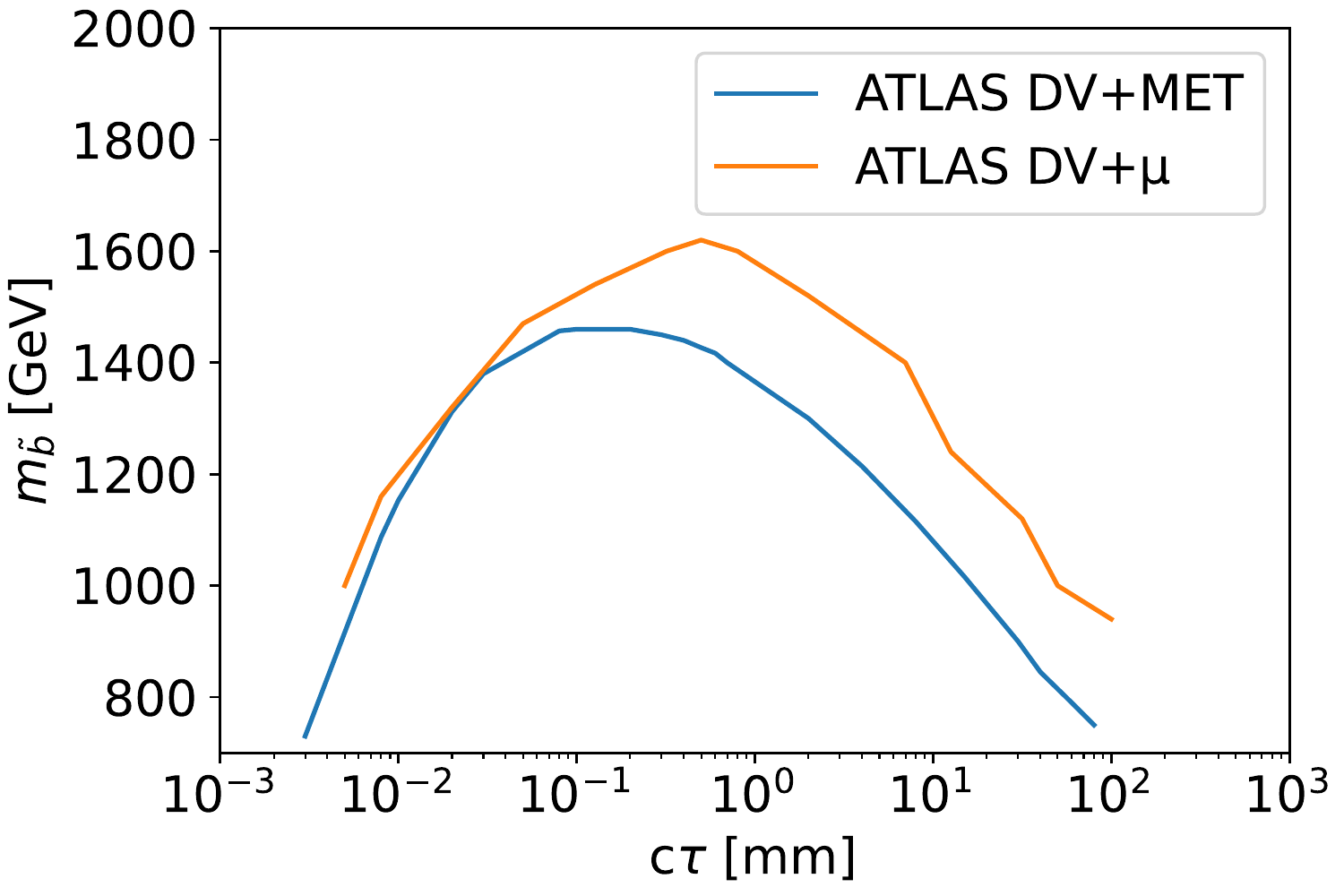}
    \includegraphics[scale=0.5]{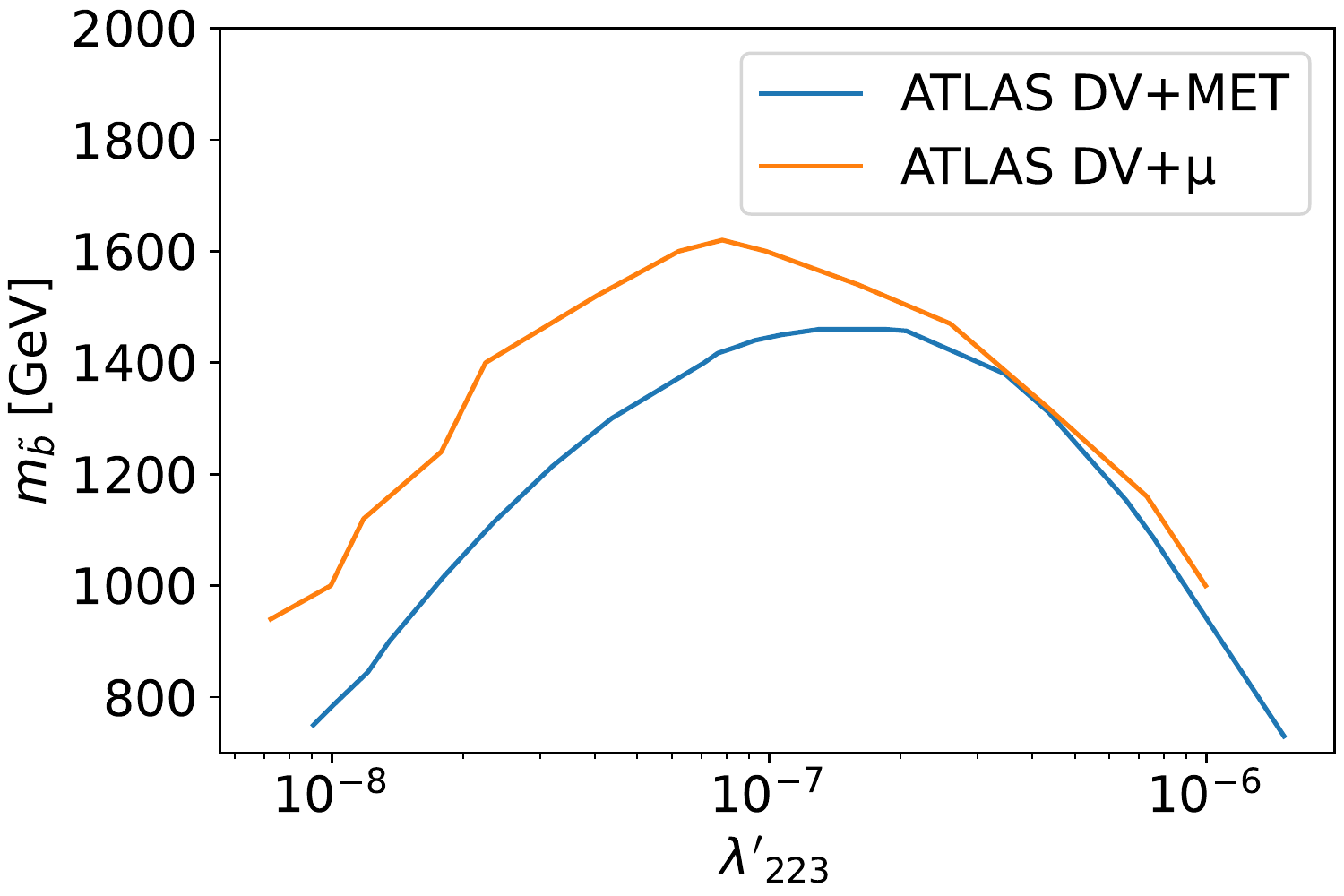}
    \caption{Limits in the mass-lifetime (left) and mass-$\lambda$ $(=\lambda^{'}_{223})$ plane of the LLP sbottoms. 
    \label{fig:strongLLP_bounds}}
\end{figure}

The limits obtained with {\tt CheckMATE} are presented in Fig.~\ref{fig:strongLLP_bounds} in the plane corresponding to the lifetime and mass of the bottom squark. We observe that, in this configuration where muon and MET productions are set equal, the DV+muon search proves slightly more competitive than the DV+MET analysis (which is consistent with the respective limits placed on the cross-sections by these searches in their respective benchmark scenario). The most constraining limits are placed on lifetimes $\sim1$~ns and exclude sbottom masses up to $\sim1.6$~TeV. This is more competitive than limits from the R-parity conserving scenario with $\tilde b \rightarrow b \tilde{\chi}^0$, which places a limit of 1270 GeV for a massless $\tilde{\chi}^0$~\cite{Aad:2021jmg}. 

Two types of prompt searches would also constrain the same model in a different parameter space searches for promptly-decaying leptoquarks, or usual sbottom limits (in the presence of some light LSP with the decay $\tilde b \rightarrow c \tilde \chi^+ \rightarrow c \bar \ell \nu \tilde \chi^0 $).  There have not been direct searches for sbottoms using this topology.  Limits on squarks in the $2\ell + 2\mathrm{jets} + \mathrm{MET}$ or in $ 2\mathrm{jets} + \mathrm{MET}$ are both expected to be much smaller than the standard topology ($\tilde b \rightarrow b \tilde \chi^0$). Limits on standard sbottom decay with full run 2 data are currently between 600-1270~GeV~\cite{Aad:2021jmg} depending on the mass of the final invisible particle. Although there have been searches for third generation leptoquarks, they focus on decays into tops~\cite{Aad:2021rrh, Sirunyan:2020zbk} (requiring e.g.\ b-tagged jets) and therefore do not apply directly to our model. 

\FloatBarrier

\section{Comments and Outlook}

In this paper, we present the implementation of a new class of analyses in the {\tt CheckMATE} package, dedicated to long-lived particle searches, which can be used to reinterpret experimental limits on new physics models. We demonstrate the interplay of these searches in detecting strongly or weakly charged LLPs and obtain limits comparable to prompt limits in certain ranges of lifetime. In addition, these searches provide a way to probe couplings that can even be much smaller than those currently observable via low-energy intensity-frontier experiments, e.g.\ by measuring meson decays.  

For our two scalar models, we find that the electroweak model can be constrained up to a mass of 480 GeV with the CMS dilepton search for lifetimes under 10 cm.  The charged track search is able to set bounds up to about 400 GeV for large lifetimes, larger  than 10 m. 

We find an important gap in the search for electroweakly charged LLPs decaying to leptons in the intermediate lifetime range.  Since the disappearing track search employs a lepton veto, it is not possible currently to understand whether  models with decays into leptons can be observed (as the leptons emerge with potentially very large impact parameters and not from the primary vertex).  It would be fruitful to both improve the experimental search criteria as well as to provide efficiency for lepton identification based on $p_T$ and $d_0$ to remedy this situation.    

For the strongly charged LLP, our limits are stronger than typical SUSY searches for particles with the same quantum numbers because of the much smaller backgrounds in these exotic searches.  An important gap here is that interaction of these particles with the detector material results in the efficiencies being dependent on some more unknown parameters than just the mass and lifetime. This is highlighted in Appendix C.  Also, this same issue prevents us from naively applying the charged track search to strongly interacting LLPs where charge exchange with the detector material becomes important.

Due to the absence of standardized detector objects in such searches, each re-interpretation heavily depends on the parametrized efficiencies published by the experiments. The validation of the five searches considered here has shown that this method gives relatively good results, which should allow the recast of the experimental limits for a very diverse range of models sharing an LLP as a common feature. However, since the efficiencies do rely on identifying the right truth-level particle via a user input and PDG code (possibly assigned ad hoc by the user for new particles), we do advise user vigilance when using these results.  

Users can implement their own versions of an LLP analysis using the {\tt AnalysisManager}, which simplifies the setting up of detector parameters, stores the expected and observed events reported by the experiment in the correct format for future use in statistical calculations and provides a skeleton C++ analysis code with access to all detector objects.  Currently, searches based on ionisation, like those for monopoles or multi-charged objects, are not possible to implement because the relevant efficiencies are not publicly available.  However, if the efficiency tables based on mass, charge and momentum were to be provided by the experiments in the future, it would be possible to implement them as well, without further difficulties.  The latest version of {\tt CheckMATE} can always be downloaded from
\url{https://github.com/CheckMATE2}.  Documentation and validation notes can be found at  \hbox{\url{https://checkmate.hepforge.org/}}.

\acknowledgments

We thank Giovanna Cottin and Andre Lessa for useful discussions.
R.~RdA acknowledges partial funding/support from the Elusives ITN (Marie Sk\l{}odowska-Curie grant agreement No 674896), the ``SOM Sabor y origen de la Materia" (FPA 2017-85985-P). 
Z.~S.~W. is supported partly by the Ministry of Science and Technology (MoST) of Taiwan with grant number MoST-109-2811-M-007-509, and partly by the Ministry of Science, ICT \& Future Planning of Korea, the Pohang City Government, and the Gyeongsangbuk-do Provincial Government through the Young Scientist Training Asia Pacific Economic Cooperation program of the Asia Pacific Center for Theoretical Physics. Z.~S.~W. also thanks the Witwatersrand University for hospitality where part of this work was initiated. KR is supported by the National Science Centre, Poland under grants: 2016/23/G/ST2/04301, 2015/18/M/ST2/00518, 2018/31/B/ST2/02283, 2019/35/B/ST2/02008, and the Norwegian Financial Mechanism for years 2014-2021,
grant DEC-2019/34/H/ST2/00707. F.D.~acknowledges support of the
BMBF Verbundprojekt 05H2018 and the DFG grant SFB CRC-110 {\em Symmetries and the Emergence of Structure in QCD}.

\appendix
\section{Usage}
In this section, we demonstrate the usage of {\tt CheckMATE} with LLP searches on an example, describing the initialization and running of a test. We assume that {\tt CheckMATE} has already been successfully installed.\footnote{For more detailed information on {\tt CheckMATE} operations, please refer to Refs.~\cite{Drees:2013wra,Dercks:2016npn}.} We then focus on a simplified SUSY scenario where only the gluino and the lightest neutralino are kinematically accessible, while the rest of the R-odd spectrum is heavy and  decoupled: we employ the same benchmark point as in section~\ref{sec:1710.04901}, with $m_{\tilde{\chi}_1^0}=0.1$\,TeV, $m_{\tilde{g}}=1.4$\,TeV, and $\tau_{\tilde{g}}=0.1$\,ns. The gluinos are produced at the LHC with $13$~TeV center-of-mass energy through strong interactions: the corresponding events are generated by calling \textsc{Pythia~8} internally. At this point, the {\tt CheckMATE} user needs to provide input to the program, either via a parameter card or via command line. We consider the minimal input through text file (parameter card) below. For the command line call, we refer the reader to Ref.~\cite{Kim:2015wza}.
\begin{bigtextfile}{checkmate\_example.in}
\begin{Verbatim}
[Parameters]
Name: ExampleRun
Analyses: atlas_1710_04901
longlivedPIDs: 1000021
invisiblePIDs: 1000022
SLHAFile: ./example_run_cards/auxiliary/1710_04901_test_gluino_tau_0.1.slha

[1710_04901_example_run]
Pythia8Process: p p > go go
MaxEvents: 10000
XSect: 0.0284 PB
\end{Verbatim}
\end{bigtextfile}
The text command file is structured in blocks,
which are introduced by expressions between brackets and contain one or more \verb@Key: Value@ pairs. 

\verb@[Parameters]@ provides general settings that are common to all processes in the {\tt CheckMATE} run. In the example above, the run is called \texttt{ExampleRun}, which also determines the name of the result directory. The \texttt{Analyses} parameter selects the list of analyses against which the generated events are tested: a list of pre-defined identifiers involving LLP searches is provided in Table~\ref{tab:all_analyses}. Only the displaced-vertex+MET search is kept in the file above. The PDG code of the LLP (gluino) is defined in the \texttt{longlivedPIDs} entry, while that of exotic stable electroweakly-interacting particles (the neutralino LSP in our case) is fed via \texttt{invisiblePIDs}.  In the current CheckMATE version, only a single \texttt{longlivedPIDs} and \texttt{invisiblePIDs} are supported. Finally, the path to the SLHA \cite{Skands:2003cj,Allanach:2008qq} file defining the spectrum is provided after the entry \texttt{SLHAFile}.

\begin{table}[tbh!]
    \centering
    \begin{tabular}{|l|l|}
    \hline
     {\bf Analysis}    &  {\bf CheckMATE identifier} \\
     \hline
      ATLAS disappearing track EW   & \verb|atlas_1710_04901_ew|\\
      ATLAS disappearing track QCD   & \verb|atlas_1710_04901_strong|\\
      ATLAS heavy charged track (EW only)   & \verb|atlas_1902_01636|\\
      ATLAS displaced vertex (DV with lepton veto) & \verb|atlas_1710_04901| \\
      ATLAS muon plus displaced vertex & \verb|atlas_2003_11956|\\
      \hline
      CMS 8 TeV displaced leptons & \verb|cms_1409_4789| \\  
      CMS 13 TeV displaced leptons & \verb|cms_pas_exo_16_022| \\
    \hline
    \end{tabular}
    \caption{Table of implemented LLP analyses and their identifiers that can be used in the CheckMATE run card.}
    \label{tab:all_analyses}
\end{table}

The following [X] blocks list individual production processes (only one in our example), with X the corresponding (and freely chosen) identifier. \texttt{Pythia8Process} defines the considered production mode. \texttt{MaxEvents} sets the number of generated events. \texttt{XSect} contains the corresponding cross-section value at $13$\,TeV center-of-mass energy, taken from the LHC cross-section Working Group at NNLO$_{\text{approx}}$+NNLL \cite{gluinoprod}.

\texttt{CheckMATE} is executed with the following command:

\begin{textfile}{Terminal}
\begin{Verbatim}[commandchars=\\\@\@]
   $CMDIR/bin: \userinputcolor ./CheckMATE checkmate\_example.in
\end{Verbatim}
\end{textfile}

The program responds with a summary of the inputted settings for the considered run and prompts confirmation from the user.
\begin{bigtextfile}{Terminal}
\begin{Verbatim}
  ____ _               _    __  __    _  _____ _____ ____  
 / ___| |__   ___  ___| | _|  \/  |  / \|_   _| ____|___ \ 
| |   | '_ \ / _ \/ __| |/ / |\/| | / _ \ | | |  _|   __) |
| |___| | | |  __/ (__|   <| |  | |/ ___ \| | | |___ / __/ 
 \____|_| |_|\___|\___|_|\_\_|  |_/_/   \_\_| |_____|_____|
\end{Verbatim}
\begin{Verbatim}[commandchars=\\\@\@]
The following settings are used:
Analyses: 
	atlas_1710_04901 ( )
E_CM: 13.0
Processes: 
	Process Name: 1710_04901_example_run
	Input Cross section: 0.0284 PB
	Associated event files and/or Monte-Carlo generation runs:
		 Pythia8 Events
			 - internal identifier:  '1710_04901_example_run'
			 - simplified SUSY process: p p > go go
			 - at most 10000 events are generated and analysed



Output Directory: 
	$CMDIR/results/ExampleRun
Additional Settings: 
	 - SLHA file path/1710_04901_test_gluino.slha will be used for event generation
	 - The following PIDs will be considered as invisible for the detector: 1000022
	 - The following PIDs will be considered as long lived for the detector: 1000021
Is this correct? (y/n) 

\end{Verbatim}
\end{bigtextfile}
In this simple example, \texttt{CheckMATE} tests against the displaced-vertex+MET search from ATLAS. 

Returning \texttt{y} launches the (\texttt{Pythia}+)\texttt{CheckMATE} run. During its course, updates of the test status are printed on the screen. Finally, the program displays the result summary:
\begin{bigtextfile}{Terminal}
\begin{Verbatim}[commandchars=\\\@\@]
Evaluating Results
Test: Calculation of approximate (fast) likelihood given in results folder
Result: Excluded
Result for r: 150.498232611
Analysis: atlas_1710_04901
SR: SR1
\end{Verbatim}
\end{bigtextfile}
In agreement with the ATLAS result presented in Ref.~\cite{Aaboud:2017iio}, or the left plot in Fig.~\ref{fig:validation_1710_04901}, the benchmark point is thus excluded ($r\gg1$) by the only analysis (\verb|atlas_1710_04901|) considered in the {\tt CheckMATE} test, in the signal region \verb|SR1|.

\subsection*{Note concerning the compatibility of LLP and prompt searches in CheckMATE}

The explicit partitioning between prompt and LLP searches is in particular motivated by the absence of in-built veto against LLP phenomena in the implemented prompt searches, leading to inconsistent results when testing events with particles having displaced production or decay points. Obviously it would be possible to conservatively test LLP spectra against prompt searches by calculating the fraction of LLPs decaying before hitting the tracker in the simulated events, then multiplying the $r$-value obtained with {\tt CheckMATE} by this fraction. Such an approach was employed in a recasting tool \textsc{SModelS}~\cite{Ambrogi:2018ujg}. Yet, such a test against prompt searches would still require a separate {\tt CheckMATE} run as compared to the test against LLP searches.  

In fact, the fundamental reason for this separation is that the standard efficiencies on reconstructed detector level objects (such as jets, leptons, missing transverse energy) used in prompt searches do not necessarily apply to the case of LLP searches. Individual LLP analyses e.g.\ employ truth-level information (i.e.\ properties of the MC event generation prior to the fast detector simulation), with recasting efficiencies provided by the experimental collaborations explicitly applying on truth-level objects.  We provide a new directory in the top-level {\tt data} directory called {\tt tables} where any efficiency tables can be stored. ROOT file format is preferred as it is easily available from experimentalists but any format that can be read directly from the analysis code in is allowed. 

There are also situations where two different LLP searches should not be run simultaneously. For example the disappearing track search has two different efficiency tables published by the ATLAS disappearing track search for tracks due to electro-weak particles and strong particles.  Only one of these should be on at a time.  In general, we suggest that users build up by hand the list of LLP analyses that they wish to run simultaneously.  To facilitate this, Table~\ref{tab:all_analyses} lists the  
{\tt CheckMATE} identifiers of all the implemented LLP analyses.

\label{app:usage}

\section{Implementation details: Displaced Lepton}
\label{app:A}

We use simulated Monte Carlo samples to evaluate the acceptance of the search regions. The samples are generated for the process  $\textnormal{pp}\rightarrow\textnormal{\~t}_1\textnormal{\~t}_1^*$, for a top squark mass of 500 (700) GeV, using \textsc{MadGraph5} \cite{Alwall:2014hca} to produce the LHE file for both the 8 TeV and 13 TeV runs. The top squark further decays via an RPV vertex into electrons and muons with a branching ratio of 0.5 each.  Total production cross section is normalized to the stop-pair production cross section at NLO.\footnote{The LHC SUSY cross section working group publishes updated calculations of total cross sections at \mbox{\url{https://twiki.cern.ch/twiki/bin/view/LHCPhysics/SUSYCrossSections}}}   To recover the assumption of lepton universality, we multiply the overall cross section with a factor of 2/3.  Events are generated using \textsc{Pythia~8}, followed by  \textsc{Delphes}  for simulation of the CMS detector acceptance and efficiencies respectively. \textsc{FastJet~3} \cite{Cacciari_2012} was used to construct jets, using the anti-kt algorithm \cite{Cacciari_2008} with a distance parameter of 0.4. Ten thousand events are generated for the described process, for each of the following proper lifetimes of the top squark: 1 mm, 10 mm, 100 mm. Additionally, for the 13 TeV analysis, a sample is generated for the stop proper lifetime 1 m.

\subsubsection*{Event Selection}
Throughout this section, the 13 TeV criteria (wherever they differ) are shown within a bracket following the 8 TeV requirements.  At the preselection stage, events with exactly one electron and one muon with opposite charge are singled out. Further conditions on the leptons to pass the detector trigger requirements are as follows: 
\begin{itemize}
    \item $p_T^\mu > 25 (40) \textnormal{ GeV}$
    \item $p_T^e > 25 (42) \textnormal{ GeV}$
    \item $|\eta^l| < 2.5 (2.4) $
\end{itemize}
Additional isolation requirements are placed on the leptons by requiring that the sum of $p_T$ of all tracks that fall within some $\Delta R$ of the lepton track are smaller than a fraction $f$ of the lepton-$p_T$. Here are the requirements in terms of $(\Delta R, f) $ for the different cases.\begin{itemize}
    \item Muons: (0.5, 0.15)
    \item Electrons with $1.56 (1.57) < |\eta^e| < 2.5 (2.4)$: (0.3, 0.065)
    \item Electrons with $|\eta^e| < 1.44$: (0.3, 0.035)
\end{itemize}
Electron candidates are rejected if they have $1.44 < |\eta| < 1.56$ due to reduced reconstruction performance of electrons in the overlap region between the barrel and endcap detectors. For the 8 TeV search, there must be no jets within a cone of $\Delta R = 0.5$ around either lepton and the electron and muon must be separated by $\Delta R > 0.5$. There is no jet-isolation requirement for the 13 TeV search.

Events passing the preselection criteria are further classified according to the impact parameters of the leptons. The impact parameter $|d_0|$ is defined as the distance of closest approach of the helical trajectory of the lepton in the transverse plane to the beam axis.
$$
|d_0^\ell| = \frac{|p_x^\ell L_y^\ell - p_y^\ell L_x^\ell|}{p_T^\ell}
$$
where $p_x^\ell$ and $p_y^\ell$ are the radial components of the lepton's momentum while $L_x^\ell$ and $L_y^\ell$ are the radial components of the decay vertex (position) of the top squark from which the lepton originates. By requiring the leptons to have a large enough impact parameter, one reduces the likelihood of the leptons of having originated from standard model processes. Both leptons are therefore required to have a minimum impact parameter $d_0^\ell > 0.1 (0.2)$ mm.  Three signal regions are defined based on these impact parameters:
\begin{itemize}
\item SR1: $d_0>0.2$~mm for both leptons and $d_0<0.5$~mm for at least one;
\item SR2: $d_0>0.5$~mm for both leptons and $d_0<1$~mm for at least one;
\item SR3: $d_0>1$~mm for both leptons.
\end{itemize}
Furthermore, events in which either lepton has an impact parameter $d_0 > 20(100)$ mm is rejected, in order to ensure that the leptons originate within the pixel layer.

\subsubsection*{Comparison}
To compare the results of this analysis with the values determined by the CMS collaboration \cite{CMS-PAS-EXO-16-022, Khachatryan_2015}, we define the percentage difference as
\begin{equation}
    \textnormal{diff} = \frac{|n_{\textnormal{CheckMATE}}-n_{\textnormal{CMS}}|}{n_{\textnormal{CMS}}}
\end{equation}{}
where $n$ is the number of events that pass all selections in each signal region.

The results as well as the comparison with the CMS analysis are shown in Tables \ref{table:1} and \ref{table:2}.

\begin{center}
\begin{table}[h]
\begin{tabular}{|r | c | c | c | c | c | c | c | c | c|}
    \hline
     $c\tau$ & \multicolumn{3}{c|}{Search Region 1} & \multicolumn{3}{c|}{Search Region 2} & \multicolumn{3}{c|}{Search Region 3}\\
     \hline
      & CM & CMS & diff & CM & CMS & diff & CM & CMS & diff\\
     \hline
     1 mm & 27.2 & 30.1 $\pm$ 5.35 & 9.63\% & 6.86 & 6.54 $\pm$ 1.21 & 4.89\% & 1.26 & 1.34 $\pm$ 0.28 & 5.97\%\\
     10 mm & 37.6 & 35.3$\pm$ 6.25 & 6.52\% & 32.0 & 30.3 $\pm$ 5.35 & 5.61\% & 50.0 & 51.3 $\pm$ 9.06 & 2.53\%\\
     100 mm & 4.78 & 4.73 $\pm$ 0.88 & 1.06\% & 6.13 & 5.57 $\pm$ 1.03 & 10.0\%& 26.5 & 26.3$\pm$ 4.65 & 0.760\%\\
     \hline
\end{tabular}    
\caption{Comparison of the number of expected events in the three search regions for the Displaced Lepton search with the CMS detector at 8 TeV with {\tt CheckMATE} (CM), for the process $pp \rightarrow \textnormal{\~t}_1 \textnormal{\~t}_1^*$, with $\textnormal{M}_{\textnormal{\~t}_1} = 500$ GeV and stop quark lifetimes of 1, 10 and 100 mm. The CMS results are presented here with statistical and systematic uncertainties combined in quadrature.}
\label{table:1}
\end{table}
\end{center}

\begin{center}
\begin{table}
\begin{tabular}{|r | c | c | c | c | c | c | c | c | c|}
    \hline
     $c\tau$ & \multicolumn{3}{c|}{Search Region 1} & \multicolumn{3}{c|}{Search Region 2} & \multicolumn{3}{c|}{Search Region 3}\\
     \hline
      & CM & CMS & diff & CM & CMS & diff & CM & CMS & diff\\
     \hline
     1 mm & 3.85 & 3.80 $\pm$ 0.2 & 1.24\% & 0.99 & 0.94 $\pm$ 0.06 & 4.8\% & 0.19 & 0.16 $\pm$ 0.02 & 20\%\\
     10 mm & 5.18 & 5.20 $\pm$ 0.4 & 0.38\% & 4.6 & 4.1 $\pm$ 0.3 & 11\% & 7.6 & 7.0 $\pm$ 0.3 & 8.0\%\\
     100 mm & 0.86 & 0.80 $\pm$ 0.1 & 7.0\% & 1.04 & 1.0 $\pm$ 0.1 & 3.97\% & 5.4 & 5.8 $\pm$ 0.2 & 7.5\%\\
     1000 mm & 0.03 & 0.009 $\pm$ 0.005 & $\sim$260\% & 0.05 & 0.03 $\pm$ 0.01 & 59\% & 0.45 & 0.27 $\pm$ 0.03 & 67\%\\
     \hline
\end{tabular}   
\caption{Comparison of the number of expected events in the three search regions for the Displaced Lepton search with the CMS detector at 13 TeV with {\tt CheckMATE} (CM), for the process $pp \rightarrow \textnormal{\~t}_1 \textnormal{\~t}_1^*$, with $\textnormal{M}_{\textnormal{\~t}_1} = 700$ GeV and stop quark lifetimes of 1, 10, 100 and 1000 mm. The CMS results are presented here with statistical and systematic uncertainties combined in quadrature.}
\label{table:2}
\end{table}
\end{center}

The analysis was conducted over top squark masses ranging from 300 (400) GeV to 850 (900) GeV, and top squark lifetimes $c\tau$ from 0.1 mm to 1 m. A simultaneous counting experiment was performed on the three bins of the three signal regions, and the resulting exclusion contours compared to those provided by CMS~\cite{CMS-PAS-EXO-16-022, Khachatryan_2015}. Figure \ref{fig:DLexclusion} of section~\ref{sec:DLS} shows these exclusion contours for the 8 TeV and 13 TeV analyses. It is observed that the {\tt CheckMATE} 8 TeV exclusion contours agree well with CMS, except for $c\tau$ between about 7 mm to 70 mm, where the {\tt CheckMATE} limits are stronger. 

For the $13$~TeV analysis, the absence of available efficiencies complicate the recast. The use of the $8$~TeV values results in satisfactory agreement at small lifetimes up to about $5$~mm. However, as the $13$~TeV search extends the allowed $d_0$ up to $100$~mm, it is necessary for the recast to assess the efficiency in the $d_0$-bin $20-100$~mm. The lepton $d_0$ efficiency is linearly extrapolated at 22 mm using the values in the range [12~mm, 20~mm] for both electrons and muons. Keeping the electron $d_0$ efficiency in the added bin constant, we vary the muon $d_0$ efficiency in the added bin and minimize the $\chi^2$ fit over all resulting signal regions for the benchmark stop lifetimes 100~mm and 1000~mm. The minimum $\chi^2$ value occurs for a muon $d_0$ efficiency of 0.01 in the added bin. Using this value, we similarly minimize the $\chi^2$ fit for electron $d_0$ efficiency values in the added bin. This minimum occurs for an electron $d_0$ efficiency of 0.06. The resulting exclusion plot still exhibits stronger limits than those of the CMS publication, but shows  a significant improvement over the exclusion plot obtained with only the 8 TeV lepton reconstruction efficiencies.

\section{Implementation details: Displaced Vertex and MET}
\label{app:C}

\subsubsection*{Monte Carlo Simulation Samples}
As a template for this search, the ATLAS collaboration considered the (strong) production of a pair of long-lived gluinos, then decaying into light quarks and a stable neutralino: $pp\to\tilde{g}\tilde{g}$, $\tilde{g}\to q \bar{q}\tilde{\chi}_1^0$. The gluino decay branching ratios are set to $50\%$ for both the $d\bar{d}\tilde{\chi}_1^0$ and $u\bar{u}\tilde{\chi}_1^0$ channels. Two values of the gluino mass are studied, $m_{\tilde{g}}=1.4$~TeV and $2$~TeV, with varying lifetimes between $\tau_{\tilde{g}}=0.003$~ns and $50$~ns. The neutralino mass is either set to $m_{\tilde{\chi}_1^0}=100$~GeV or kept at $m_{\tilde{g}}-100$~GeV. We generate events internally to {\tt CheckMATE}, via the \textsc{Pythia~8} interface, also carrying out the hadronization. Jets are reconstructed with \textsc{FastJet~3} \cite{Cacciari_2012}, using the anti-kt algorithm \cite{Cacciari_2008}, with a distance parameter of $0.4$. The cross-section is varied by rescaling the number of events by the ratio (imposed cross-section)/(SUSY cross section), so that the limits are independent of the actual SUSY production cross-sections (or their LO calculation). Samples of $10^6$ events are considered for the relevant mass, lifetime and cross-section values.

While we principally consider unmatched events, we will also discuss the impact of jet radiation on individual scenarios. For this, we employ \textsc{MadGraph5 2.7.0 } \cite{Alwall:2014hca} and add up to two jets using the MLM method with xqcut parameter of $100$~GeV.

\subsubsection*{Event selection}
The event selection follows the recast instructions provided by the ATLAS collaboration \cite{DVrecast}. It includes a preselection defining the event- and vertex-level acceptances for truth-level particles, and then the application of parameterized efficiencies. This recast strategy has been successfully applied in the past (see Contribution~$22$ in \cite{Brooijmans:2018xbu} and the publicly available codes \cite{Cottincode,Lessacode}).

\subsubsection*{Preselection}
The missing energy $E_{\text{T,true}}^{\text{miss}}$ is defined at truth-level as the vector sum of the momenta of the stable invisible weakly-interacting particles ($\tilde{\chi}_1^0$ and neutrinos in the considered benchmark scenario). It is requested to satisfy the cut: $E_{\text{T,true}}^{\text{miss}}>200$~GeV.

Jet requirements are applied on $75\%$ of the events. These demand either one truth jet with $p_\text{T}>70$~GeV or at least two jets with $p_{\text{T}}>25$~GeV, satisfying in both cases a trackless requirement: the scalar sum of the charged particle $p_{\text{T}}$ in the jet should not exceed $5$~GeV for particles with small impact parameter ($d_0$). We interpret the small impact-parameter condition as $d_0<2$~mm.

DVs are reconstructed from stable charged particles. The event should contain at least one DV in the fiducial volume: the transverse distance from the interaction point should lie between $4~\text{mm}$ and $300~\text{mm}$, as well as $|z|<300~\text{mm}$. The DVs should contain at least $5$ selected decay products, i.e.~stable and charged particles with $p_{\text{T}}>1$~GeV and an approximate transverse impact parameter $d_0>2$~mm. The condition on the $p_{\text{T}}$ of the selected decay product assumes a charge $|q|=1$, which we regard as implicitly fulfilled.

\subsubsection*{Efficiencies}
Efficiencies of two origins are considered. The first one corrects $E_{\text{T,true}}^{\text{miss}}$. The second one applies on the reconstruction of the DV. Both are provided by the experimental collaboration \cite{DVdata}.

The results presented in Fig.~\ref{fig:validation_1710_04901} seem to indicate a satisfactory performance of the recast strategy in scenarios involving a sizable mass-splitting between the LLP and the LSP. In addition, we were able to largely recover the cutflow information \cite{DVrecast} (as far as a comparison between truth- and detector-level events is possible). 

\subsubsection*{Observed inconsistencies in provided efficiency parametrisation}
\begin{figure}[tbh]
\begin{center}
    \includegraphics[width=0.49\textwidth]{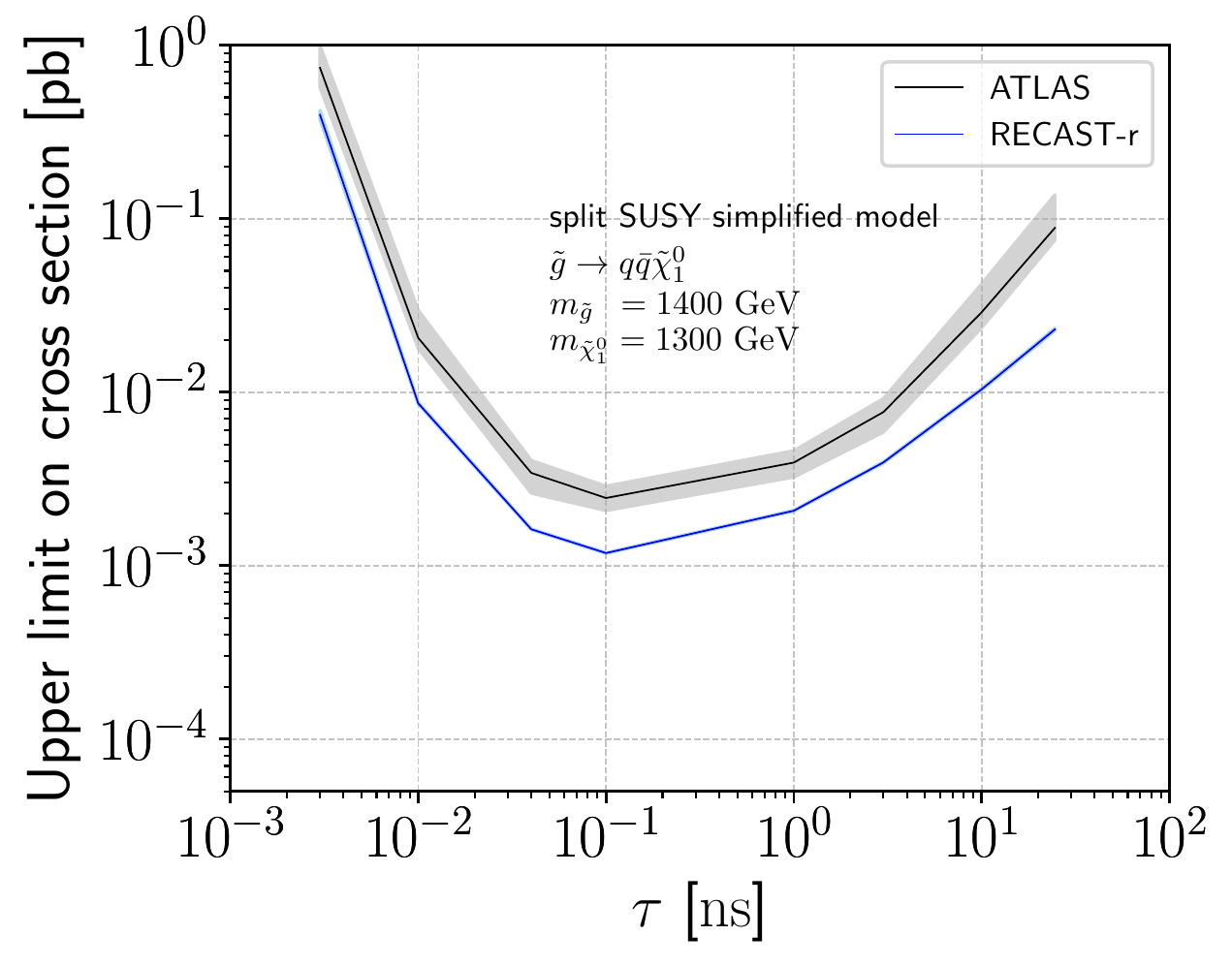}
	\includegraphics[width=0.49\textwidth]{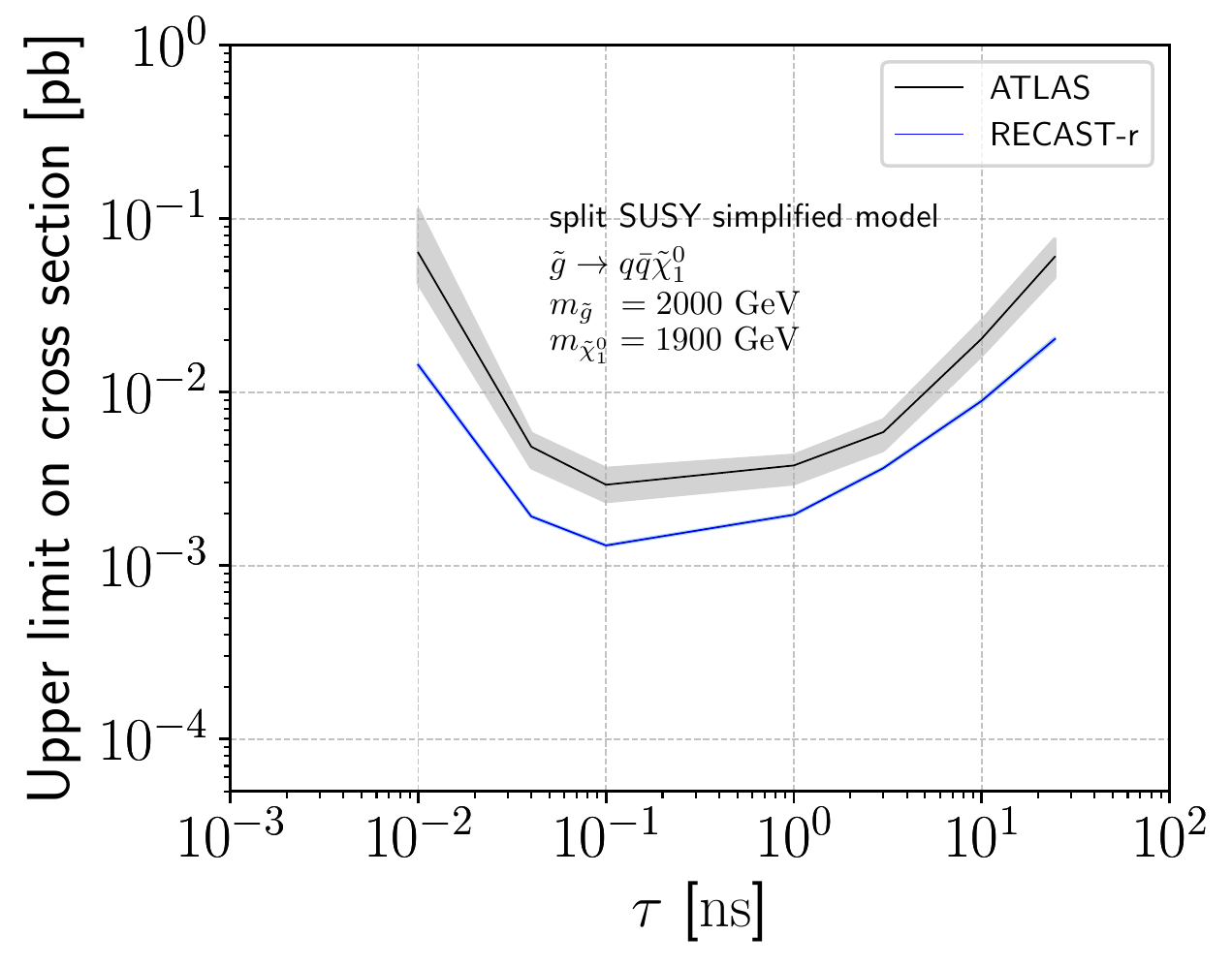}
	\includegraphics[width=0.49\textwidth]{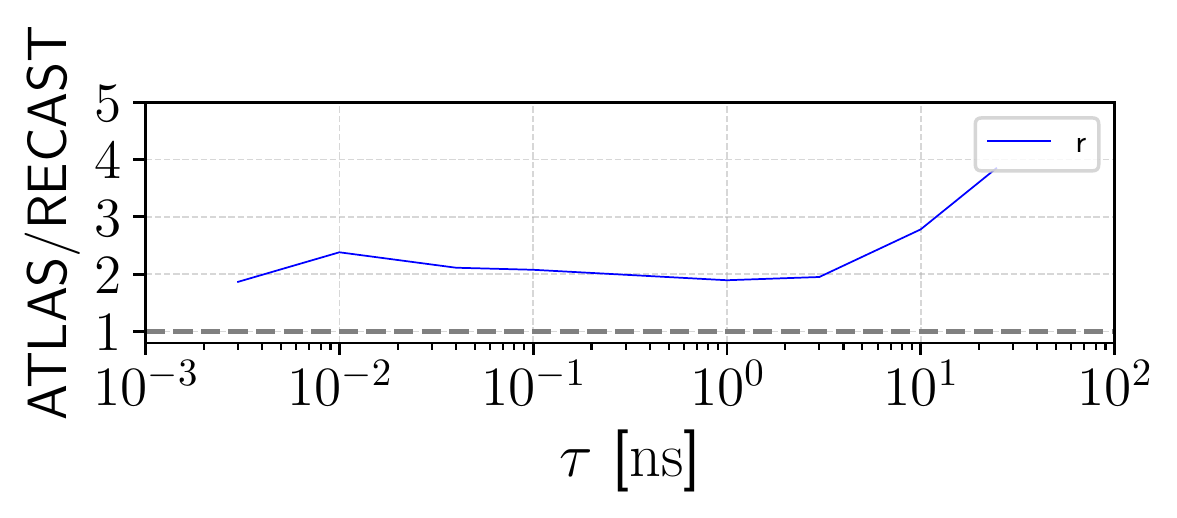}
	\includegraphics[width=0.49\textwidth]{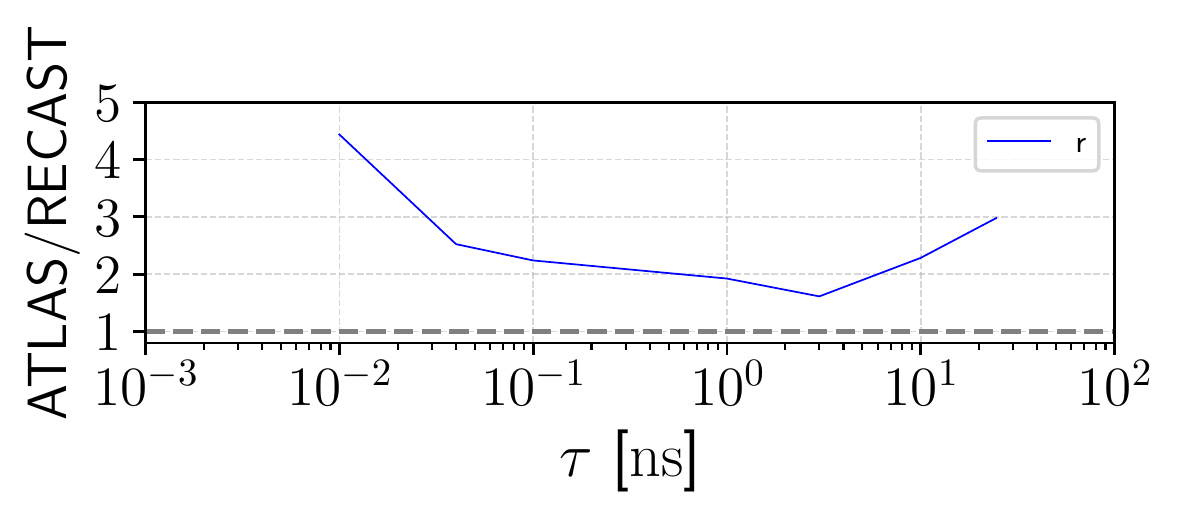}
\end{center}
	\caption{Scenario with compressed spectrum -- validation plots. Same conventions as in Fig.~\ref{fig:validation_1710_04901}. We only present the results obtained with the simplified {\sc CheckMATE} criterion.}
	\label{fig:validation_1710_04901_compressed}
\end{figure}

We then test the scenario with compressed SUSY spectra, i.e.~a $100$~GeV mass-splitting between gluino and the neutralino: see Fig.~\ref{fig:validation_1710_04901_compressed}. In this case, the limiting cross-sections obtained with the recast information differ from the experimental observed limits by a sizeable deviation: on average, a factor $2$ to $3$.
This issue appears to be generic with this recast strategy, as was confirmed by the authors of \cite{Cottincode,Lessacode}. The impact of R-hadronization on the definition of the missing transverse energy was put forward as a possible factor. We also note that the perspectives for 
implementing a successful recast look bleak in a region where 
cross-section limits show a strong variation with the kinematics, such 
as the compressed regime of the DV+MET search (see Fig.~10a of 
\cite{Aaboud:2017iio}), due to the difficulty of matching simulation 
objects with their collider counterparts. In any case, the poor performance of the recast in this compressed configuration hints at the possible failure of the approach with parametrized efficiencies to simultaneously address all relevant spectra.  As a possible way to mitigate incomplete efficiency parametrisations, the LHC Reinterpretation Forum white paper~\cite{LHCReinterpretationForum:2020xtr} and the LLP white paper~\cite{Alimena:2019zri} recommend using multiple topologies or multiple mass benchmarks for unweighting.

\subsubsection*{Matching events}
Finally, we discuss the impact of jet radiation (matched events) for a gluino lifetime of $1$~ns. The results are collected in Table~\ref{table:matched}, allowing for up to two radiated jets. We observe that the radiation of jets affects the limiting cross-section by only a few percent in the scenario with large gluino-neutralino mass-splitting, and by $\sim10\%$ in the compressed case. Matching events thus leads to a minor effect in the considered search, justifying our focusing on unmatched events in Figs.~\ref{fig:validation_1710_04901} and \ref{fig:validation_1710_04901_compressed}. We note that the added jets affect the limiting cross sections in opposite directions for the two scenarios, relaxing the bound in the uncompressed case, while tightening it in the other case. This can be qualitatively understood as, in the compressed case, the radiation of additional jets provides possibly more energetic candidates for the jet cuts, while the jets originating in gluino decays have reduced energy due to the narrow phase space. On the contrary, in the scenario with large mass-splitting, the jets originating in gluino decays are already energetic, so that radiating further jets decreases their ability to withstand the jet cuts.

\begin {table}[th]
\begin{center}
\begin{tabular}{| c| c ||c | c|}    
\multicolumn{4}{l}{Uncompressed scenario: $m_{\tilde{\chi}_1^0}=100$~GeV.}\\\hline
\multicolumn{2}{|c||}{$m_{\tilde{g}}=1.4$~TeV} &
\multicolumn{2}{c|}{$m_{\tilde{g}}=2$~TeV}\\
  unmatched & matched & unmatched & matched \\ \hline
  $\sigma^{\text{rec}}_{\text{lim}}=0.272$ & $\sigma^{\text{rec}}_{\text{lim}}=0.281$ & $\sigma^{\text{rec}}_{\text{lim}}=0.220$ & $\sigma^{\text{rec}}_{\text{lim}}=0.228$\\ \hline
  \multicolumn{4}{l}{}\\
\multicolumn{4}{l}{Compressed scenario: $m_{\tilde{\chi}_1^0}=m_{\tilde{g}}-100$~GeV.}\\\hline
\multicolumn{2}{|c||}{$m_{\tilde{g}}=1.4$~TeV} &
\multicolumn{2}{c|}{$m_{\tilde{g}}=2$~TeV}\\
  unmatched & matched & unmatched & matched \\ \hline
  $\sigma^{\text{rec}}_{\text{lim}}=2.07$ & $\sigma^{\text{rec}}_{\text{lim}}=1.81$ & $\sigma^{\text{rec}}_{\text{lim}}=1.96$ & $\sigma^{\text{rec}}_{\text{lim}}=1.76$\\ \hline
\end{tabular}
\caption{Impact of matched events on the DV search for a gluino lifetime of $1$~ns. Matched events allow for up to two radiated jets. $\sigma^{\text{rec}}_{\text{lim}}$ represents the upper bound on the cross-section in fb, obtained from the recast procedure.}
\label{table:matched}
\end{center}
\end {table}

\FloatBarrier

\section{Implementation details: Displaced vertex and a muon\label{app:E}}
\subsection*{Monte Carlo Samples}
As a benchmark process for the search involving displaced vertices and displaced muons~\cite{Aad:2020srt} the stop pair production was considered, followed by the RPV decay: $pp \to \tilde{t}_1 \tilde{t}_1$, $\tilde{t}_1 \to \mu\, q$. The Monte Carlo samples for cutflows and exclusion limits were generated using {\sc MadGraph 2.7.2}~\cite{Alwall:2014hca} with up to two additional partons matched using CKKW-L procedure~\cite{Lonnblad:2011xx}. Showering was performed by \textsc{Pythia~8.244}~\cite{Sj_strand_2015}. To minimise the impact of statistical uncertainty the sample for each grid-point or cutflow scenario were 10--20 times larger than the nominal number of events. The samples were normalized to the approximate next-to-next-to-leading order in the strong coupling constant with a soft gluon resummation (approximate NNLO+NNLL)~\cite{Beenakker:2016lwe,Beenakker:1997ut,Beenakker:2010nq,Beenakker:2016gmf}.

\subsection*{Event selection and signal regions}
The event selection defines two mutually exclusive trigger-based signal regions: $E_\mathrm{T}^\text{miss}$ Trigger SR and Muon Trigger SR. The first SR should be selected by the MET trigger and requires $E_\mathrm{T}^\text{miss}> 180$~GeV and a muon with $p_T > 25$~GeV, while the second one should be selected by the muon trigger with muon $p_T > 60$~GeV  and $E_\mathrm{T}^\text{miss} <  180$~GeV. For both signal regions the muon vertex should be at least 2 mm away from the primary vertex. The selection also requires a displaced vertex at least 4 mm away from the primary vertex, with at least 3 associated tracks and its visible invariant mass calculated from the four-momenta of the associated tracks $m_\text{DV}> 20$~GeV (assuming each track originates from a charged pion).    

Displaced vertices are reconstructed internally by {\tt CheckMATE} using a \texttt{DVfinder} class. Firstly, charged stable particles with production vertex separated from the primary vertex are searched for. These are then merged into vertices based on the relative position, if the distance is smaller than 1~mm. The position of the combined vertex is calculated as a weighted (with invariant momentum) average of the positions of input vertices.  The procedure continues until no further vertices can be combined and the remaining single track vertices are dropped.  

\subsection*{Validation}
For the reconstruction of events at the {\tt CheckMATE} level no specific truth particles efficiencies were used (these are available on \textsc{HEPdata}). The ATLAS analysis rejects vertices from interaction with dense detector material. The veto is imposed based on a three dimensional map of the detector and rejects $42\%$ of the events. Because the map was not public when the {\tt CheckMATE} analysis was implemented, the veto is applied as a flat rejection probability on each reconstructed displaced vertex. Due to inhomogeneous distribution of vetoed points this can potentially result in a bias for certain lifetimes compared to the experimental analysis.

The validation procedure was performed both in terms of the exclusion contour for the benchmark model, see Figure~\ref{fig:validation_2003.11958}, and of the cutflows for three different parameter points. As already noted in Section~\ref{sec:dvplusmu}, generally a good agreement is observed for the exclusion contour, except a range of lifetimes $0.01$--$0.1$~ns where the recast exclusion is significantly weaker, though still within the 2-sigma band. In order to get a better insight, in Table~\ref{tab:validation_2003} we also compare cutflows published by the ATLAS collaboration for $m_{\tilde{t}}=1.4$~TeV and three lifetimes. 

For the longest lifetime, 1~ns, the number of reconstructed events within {\tt CheckMATE} is by $40\%$ higher than that of ATLAS. This corresponds to the slightly stronger recast exclusion observed in Figure~\ref{fig:validation_2003.11958} for lifetimes $\sim$0.2--6 ns. For the intermediate lifetime, 0.1~ns, the number of {\tt CheckMATE} reconstructed events is half of the ATLAS number. This corresponds to a weaker exclusion observed in Figure~\ref{fig:validation_2003.11958} for lifetimes $ < 0.2$~ns. We note however that this falls inside the expected 2-$\sigma$ exclusion limit calculated by the collaboration. Finally, for the shortest lifetime, 0.01~ns, we observe almost a 3-fold difference. However, this region is close to the experimental sensitivity limit, as can be seen in Figure~\ref{fig:validation_2003.11958}, where the exclusion line becomes almost vertical. For this reason a poor agreement between the full simulation and recasting is not surprising. On the other hand, it also has a minimal impact on the exclusion contour.   

\begin{table}[h!] 
 \renewcommand*{\arraystretch}{1.2}
 \begin{tabular}{l|p{1.3cm}|p{1.3cm}|p{1.3cm}|p{1.3cm}|p{1.3cm}|p{1.3cm}} \toprule
                                                 & \multicolumn{2}{c|}{$\tau = 0.01$ ns} & \multicolumn{2}{c|}{$\tau = 0.1$ ns} & \multicolumn{2}{c}{$\tau = 1$ ns}\\
  Selection                                      & ATLAS & {\sc CM} & ATLAS & {\sc CM} & ATLAS & {\sc CM}      \\ \midrule
  All                                            & 64.2  &  64.2 & 64.2 & 64.2 & 64.2 & 64.2 \\
  $E_T^\mathrm{miss}$ trigger                    & 63.0  &  63.2 & 63.0 & 63.3 & 62.7 & 63.2 \\
  $E_T^\mathrm{miss} > 180$ GeV                  & 60.9  &  61.2 & 61.0 & 61.3 & 60.6 & 61.0 \\
 $\geq 1\mu$; $p_T>25$ GeV, $|\eta| < 2.5$ & 57.8  &  60.8 & 58.7 & 61.0 & 52.5 & 60.6 \\
  $2 < |d_0(\mu)| < 300$ mm                      & 11.3  &  12.8 & 49.1 & 52.5 & 49.5 & 59.2 \\
  $|z_0(\mu)| < 500$ mm                          & 11.3  &  12.8 & 49.1 & 52.5 & 49.3 & 58.1 \\
  Fake/HF/cosmic veto       & 9.1   &  9.9  & 40.0 & 42.4 & 39.4 & 48.1 \\
  At least one DV                                & 8.5   &  4.4  & 37.6 & 29.8 & 32.6 & 39.6 \\
  DV fiducial volume                             & 8.4   &  3.7  & 37.1 & 29.1 & 31.2 & 32.7 \\
  Material veto                                  & 5.3   &  2.2  & 31.0 & 16.9 & 22.2 & 19.0 \\
  $n_\text{tracks}^\text{DV} \geq 3$             & 3.8   &  1.8  & 26.0 & 15.5 & 13.7 & 17.3 \\
  $m_\text{DV} > 20$ GeV                         & 3.4   &  1.2  & 22.7 & 11.9 & 10.3 & 14.0 \\
 \bottomrule
 \end{tabular}
 \caption{Cutflow for the displaced vertex + muon search in the missing transverse energy SR. \label{tab:validation_2003}}
\end{table}

\section{Implementation details: Heavy Charged Particle Track}\label{app:heavychargedparticle}

\subsubsection*{Monte Carlo Simulation Samples}

In order to validate our implementation of this search in {\tt CheckMATE} we simulated the production of pairs of charginos and staus with \textsc{MG5\_aMC@NLO} (version 2.6.6) \cite{Alwall:2014hca} including two additional partons at leading order in combination with \textsc{Pythia~8} \cite{Sj_strand_2015}. Matching was performed using the MLM scheme \cite{Mangano:2006rw}. The resulting events in \textsc{HepMC} format \cite{Dobbs:2001ck} were processed by {\tt CheckMATE} which uses \textsc{Delphes} for the simulation of the ATLAS detector.   

The reference benchmarks used for the validation are provided in SLHA format by the ATLAS collaboration in the \textsc{HEPData} repository \cite{Maguire:2017ypu}. The pair production of charginos is based on a mAMSB \cite{Randall:1998uk} scenario while the staus production on a GMSB \cite{Giudice:1998bp} scenario. For reproducing the ATLAS results we have chosen a set of chargino and stau masses and simulated $2\times10^4$ events for each of the benchmarks considered. 



\subsubsection*{Preselection}

The following event preselection is applied:

\begin{itemize}

\item Full-detector candidates have to be electrically single-charged particles and pass the full ATLAS detector before decaying (R $>$ 12.0 m, $|z| >$ 23.0 m).

\item Events with $E_{\text{T}}^{\text{miss}} > $ 300 GeV are accepted, otherwise a trigger efficiency is applied. 

\item Single-muon trigger objects have to reach the ATLAS muon spectrometer (stable within R $>$ 12.0 m, $|z| > 23.0$ m), have a $\beta > 0.2$, a pseudorapidity $ |\eta| < 2.5$ and a minimum transverse momentum of 26 GeV.
In addition a muon trigger efficiency which is a function of $\beta$ and $|\eta|$ of the particle is applied.

\end{itemize}
Maps of trigger efficiencies are provided by the ATLAS collaboration in the \textsc{HEPData} repository \cite{hepdata}.

\subsubsection*{Search Regions}
\label{app:D}

The searches for pair-produced long-lived staus and charginos are based on a \loose and on a \tight candidate selection. The main difference between them is that the \loose selection requires two candidates while the \tight one only one. In both cases the candidates are required to have $p_T > 70$ GeV and while for the \loose selection $p > 100$ GeV is imposed, the \tight one requires $p > 200$ GeV. In addition a tighter pseudorapidity $|\eta|$ $<$ 1.65 for the \tight selection is required. Finally trigger efficiencies for \loose and \tight candidates in function of $\beta$ and $|\eta|$ of the candidates are applied. 

The signal regions are based on cuts on the \loose and \tight selections plus applying inclusive cuts on the time of flight (ToF) measurements of the particle mass $m_\mathrm{ToF}$. The reconstructed $m_\mathrm{ToF}$ is given as a function of the truth mass for the candidates and is sampled from a Gaussian using the mean and width from the respective bin in the truth mass. For candidates passing the \loose cuts two masses have to be sampled for a given truth mass and the lower of the two is used for the final counting. For \tight candidates only one mass is sampled using the respective mean and resolution. The tables parametrizing the reconstructed $m_\mathrm{ToF}$ are provided in the \textsc{HEPData} repository \cite{hepdata}.

Following the prescription above, the signal regions are defined according to the \loose and \tight cuts plus a condition on the minimum reconstructed $m_\mathrm{ToF}$ as summarized in Table \ref{table:srs}. The results of the validation are presented in Fig.~\ref{fig:validation_1902} and establish the reliability of this recast search in {\tt CheckMATE}.

\begin {table}[tbh!]
\begin{center}
\begin{tabular}{|c| c c c c|}
    \hline    
    \multicolumn{5}{|c|}{$m^\mathrm{min}_\mathrm{ToF}$ [GeV]} \\
    \hline
      \tight & 175 & 375 & 600 & 825 \\
      \loose & 150 & 350 & 575 & 800 \\
    \hline
\end{tabular}
\caption{Definition of the signal regions implemented in {\tt CheckMATE}.}
\label{table:srs}
\end{center}
\end {table}

\section{Implementation details: Disappearing Track}
\label{app:B}

\subsubsection*{Monte Carlo Simulation Samples}
We generated the SUSY signal samples in the framework of minimal AMSB model \cite{Randall:1999ee} assuming $\tan\beta=5$, with a positive higgsino mixing mass parameter, while the scalar masses are decoupled by setting $m_0=5$ TeV. The chosen benchmark contains a chargino with a 0.2 ns lifetime and a mass of 400 GeV. For the strong production channel, we select a gluino mass of 1600 GeV, a chargino mass of 500 GeV and a proper lifetime of 0.2 ns. The SUSY mass spectrum files are  generated with \textsc{ISASUSY 7.80} \cite{Baer:1993ae}. The gluino decays via off-shell squarks into the following decay channels,
\begin{equation}
  \tilde g\rightarrow q \bar{q}\tilde\chi^0_1,\quad q \bar{q}^\prime \tilde\chi^-_1,\quad q\bar{q}^\prime\tilde\chi^+_1.
\end{equation}
assuming only first and second generation partons intervene. The wino--like chargino LLP decays into a charged pion and a neutralino LSP.

The MC events for the signal are generated together with two additional radiated partons in the hard process with \textsc{MadGraph5 2.6.1}\cite{Alwall:2014hca}. The parton level events were showered and hadronized with \textsc{Pythia~8.230} \cite{Sj_strand_2015}. We used the \textsc{NNPDF 2.3LO} parton distribution function \cite{Ball:2012cx}. Renormalisation and factorisation scales were kept at the default scale of \textsc{MadGraph5}. The combination of the parton shower and the matrix elements partons was performed in the CKKW-L merging scheme \cite{Lonnblad:2001iq} and the merging scale was set to a quarter of the wino mass for the electroweak production channel or a quarter of the gluino mass for the strong production channel. The cross sections for electroweak production are calculated at NLO using \textsc{Prospino2} \cite{Beenakker:1999xh} and the normalisation for the strong production is determined at NLO+NLL accuracy with \textsc{NLLfast} \cite{Beenakker:1996ch,Kulesza:2008jb,Kulesza:2009kq,Beenakker:2009ha,Beenakker:2011fu}.

\subsubsection*{Preselection}
We apply the object removal procedure given in Ref.~\cite{Aad:2019vnb}. Electron and muon candidates must satisfy $p_T>10$ GeV and $|\eta|<2.47$ and $2.7$, respectively. The isolation criteria for final state leptons are given by the requirement that the scalar sum of the transverse momentum of tracks inside variable-size cone around leptons is less than 15$\%$ of lepton transverse momentum while the cone radius is defined as min($\Delta R=10$ GeV$/p_T$, $\Delta R=0.2\,(0.3)$) for electrons (muons). Jets are reconstructed with anti--k$_{\rm{T}}$ algorithm and $\Delta R=0.4$. Jet candidates must pass $p_T>20$ GeV and $|\eta|<2.4$. The missing transverse momentum is given as the negative vector sum of all reconstructed detector level objects, i.e.\ electrons, muons, photons and jets.

\subsubsection*{Search Regions}
For the electroweak and strong signal region, a set of common kinematic preselection cuts are demanded and for both signal regions we require a lepton veto, i.e. no isolated electrons and muons are allowed in the event. Two signal regions targeting the electroweak production channel (EW SR) and strong production channel via gluinos (strong SR) are considered. We apply parton level cuts on the tracklet (i.e.\ the long lived chargino). We demand isolated charginos with $p_T>100$ GeV. Moreover, we require geometric acceptance cuts $0.1<|\eta|<1.9$. For the EW SR, the events must have at least one jet with $p_T>140$ GeV and $E_{\text{T}}^{\text{miss}}>140$ GeV (90 GeV$<E_{\text{T}}^{\text{miss}}<$140 GeV) in the high-- (low--) missing transverse momentum region. Multijet background suppression is achieved by a $\Delta\Phi$ cut. Here, the difference in azimuthal angle between the missing transverse momentum and each of the up to four highest--$p_T$ jets with $p_T>50$ GeV is required to be larger than 1.0.

The strong SR demands that events have a jet with $p_T>100$ GeV, at least two additional jets with $p_T>50$ GeV and $E_{\text{T}}^{\text{miss}}>150$ GeV (100 GeV $<E_{\text{T}}^{\text{miss}}<$ 150 GeV) in the high--(low--) missing transverse momentum. The $\Delta\Phi$ between the missing transverse momentum and each of the up to four leading jets with $p_T>50$ GeV is required to be larger than 0.4.

\bibliographystyle{JHEP}
\bibliography{main}

\end{document}